\documentclass[12pt]{article}

\usepackage{amsmath,amssymb,mathtools,mathrsfs} 
\usepackage[italicdiff]{physics}                
\usepackage{type1cm}
\usepackage[T1]{fontenc}
\usepackage{fancyhdr}
\usepackage{appendix}
\usepackage{titlesec}
\usepackage{cite}
\usepackage{color}
\usepackage[top=25truemm,bottom=25truemm,left=20truemm,right=20truemm]{geometry}
\usepackage[dvipdfmx]{graphicx}
\usepackage{color}
\allowdisplaybreaks[4]

\baselineskip=\normalbaselineskip

\makeatletter
 
 \@addtoreset{equation}{section}
\makeatother

\begin{document}
\setcounter{footnote}{0}
\setcounter{tocdepth}{3}
\bigskip
\def\thefootnote{\arabic{footnote}}

\begin{titlepage}
\renewcommand{\thefootnote}{\fnsymbol{footnote}}
\begin{normalsize}
\begin{flushright}
\begin{tabular}{l}
UTHEP-817
\end{tabular}
\end{flushright}
  \end{normalsize}

~~\\

\vspace*{0cm}
    \begin{Large}
       \begin{center}
         {
Localization of the BFSS matrix model and \\ three-point amplitude in M-theory}
       \end{center}
    \end{Large}

\vspace{0.7cm}

\begin{center}
Yuhma A\textsc{sano}$^{1),2)}$\footnote[1]
            {
e-mail address : 
asano@het.ph.tsukuba.ac.jp},
Goro I\textsc{shiki}$^{1),2)}$\footnote[2]
            {
e-mail address : 
ishiki@het.ph.tsukuba.ac.jp},
Yoshua M\textsc{urayama}$^{1)}$\footnote[3]
            {
e-mail address : 
murayama@het.ph.tsukuba.ac.jp}

\vspace{0.7cm}

     $^{ 1)}$ {\it Graduate School of Science and Technology, University of Tsukuba, }\\
               {\it Tsukuba, Ibaraki 305-8571, Japan}\\

     $^{ 2)}$ {\it Tomonaga Center for the History of the Universe, University of Tsukuba, }\\
               {\it Tsukuba, Ibaraki 305-8571, Japan}\\
               \end{center}

\vspace{0.5cm}

\begin{abstract}
\noindent
We apply the localization method to the BFSS matrix model with a 
particular class of boundary 
conditions, that is related to a scattering problem of 11-dimensional 
M-theory. 
For the boundary condition that corresponds 
to the three-point amplitude of gravitons, we
exactly compute the partition function of the model based on the localization method.
We find that the result correctly reproduces the expected 
momentum dependence of the three point amplitude.
\end{abstract}

\end{titlepage}

\tableofcontents

\section{Introduction}

The BFSS matrix model is conjectured to give a nonperturbative 
formulation of 11-dimensional M-theory \cite{Banks:1996vh}. 
The model does not preserve all the symmetries of M-theory, 
likewise the lattice formulation of gauge theories in which 
the rotational and translational symmetries are broken to 
the discrete ones.
In particular, symmetries associated to the light-cone directions
are not visible in the matrix model.
Thus, it is very important to investigate 
whether the model can reproduce physics of M-theory 
associated to those invisible directions \cite{Polchinski:1997pz}.

In this context, a recent development 
has been provided by the paper by Herderschee and
Maldacena \cite{Herderschee:2023pza}\footnote{
See also \cite{Herderschee:2023bnc, Laurenzano:2025ywy} for  
 the soft theorem in the matrix model.}. 
They considered graviton three-point amplitude
with momentum transfer in the light-cone direction and showed that 
it agrees with the partition function of the matrix model 
with an appropriate boundary condition, which is considered to realize 
the same setup on the corresponding gravity side.
In this work, they compactified one of the spatial directions and mapped 
the BFSS matrix model to a 2-dimensional matrix model 
\cite{Dijkgraaf:1997vv} via T-duality \cite{Taylor:1996ik}.
By using various techniques (the correspondence between the open and closed channels, 
earlier results on the index computation and the presence of
 mass gap of $SU(N)$ modes),
they completed the calculation of the partition function.

In contrast, in this paper, we propose a method of 
direct calculation of the scattering amplitude 
in the BFSS matrix model based on the localization method \cite{Pestun:2007rz}.
In particular, for the boundary condition corresponding to the 
three-point amplitude, we demonstrate that 
at least when the matrix size is $2$, the partition function
has the same momentum dependence as the 
three point amplitude in M-theory.

In our method, we first give a formulation of the matrix model 
defined on a line segment, which keeps at least one off-shell supersymmetry.
The symmetry can be enhanced,
depending on boundary conditions at the edges of the line segment. 
For the boundary condition corresponding to the three-point function, 
in particular, our formulation can preserve 4 dynamical supersymmetries 
(1/4 BPS).

We then perform the localization by using one of the 4 supersymmetries
and show that the partition function is localized to solutions to the 
Nahm equation \cite{Nahm:1979yw}.
Since there is no known method of solving the Nahm equation for 
an arbitrary boundary condition or for a general matrix size, 
we restrict ourselves to the case of $N=2$ in this paper, for which 
a unique exact solution is known \cite{Braden:2022ljy}.
We complete the calculation for this case and show the agreement of the momentum dependence.

At this stage, our setup is limited to the small-$N$ case
because of the technical difficulty and thus we consider 
the finite $N$ version of the correspondence to the discrete light-cone 
quantization (DLCQ) of M-theory
\cite{Susskind:1997cw, Sen:1997we, Seiberg:1997ad}.
In this correspondence, one of the light-cone directions is compactified
with radius $R$, which roughly corresponds to the coupling
constant of the matrix model.
Since our computation is exact and valid for the strong coupling limit
(large-$R$ limit), we then compare our result with
the amplitude on the flat 11-dimensional space with small light-cone momenta\footnote{
In this limit, the light-cone momentum $p^i_- = N_i/R$ goes to
zero, but this only affects the overall constant of the three-point
amplitude and does not change the momentum dependence
studied in this paper.}. 
The amplitude in the DLCQ formalism is in general different from the
flat-space amplitude. However, we assume that this is not the case for 
the three-point amplitude, since the result in
\cite{Herderschee:2023pza} suggests that the amplitude takes the 
same form even when the matrix size is finite.

One advantage of our method is that it enables direct and exact 
computation of the BFSS model without relying on conjectured dualities. 
Furthermore, our method is applicable in principle to more general settings, 
such as general matrix sizes and general boundary conditions 
corresponding to higher point amplitudes.

This paper is organized as follows.
In section \ref{section2}, we review the three-point amplitude in 11-dimensional M-theory.
In section \ref{section3}, we give a supersymmetric 
formulation of the BFSS matrix model on a line segment.
In section \ref{section4}, we perform the localization and obtain the final result.
Section \ref{section5} is devoted to summary and discussion.

\section{Three-point amplitude in M-theory}
\label{section2}

Let us review the three point amplitude of 11-dimensional 
gravitons \cite{Herderschee:2023pza}
with momenta $p^i_\mu$ with $i=1,2,3$ satisfying 
the on-shell condition $(p^i_\mu)^2=0$, as well as 
the conservation law, $\sum_{i=1}^3 p^i_\mu =0$. 
It is easy to see that for real momenta in the Lorentzian signature, 
these conditions are satisfied 
only when the three momenta are all proportional to each other. 
In order to study the analytic structure of the three point amplitude,
it is useful to consider complexified momenta, or alternatively, 
to consider $(2, 9)$ signature, where one of the space direction
is Wick-rotated to the negative signature.

We consider the latter prescription and 
set the metric of 11-dimensional flat space with $(2, 9)$-signature as
\begin{align}
ds^2 &= -2dx^+ dx^-  -dy^2 +dx^2 + \cdots,
\nonumber\\
&=-2dx^+ dx^- +dX d\Phi + \cdots,
\label{11d metric}
\end{align}
where $x^\pm$ are the light-cone coordinates,
$y$ is the coordinate for the
Wick-rotated spatial direction, and the rest part 
$dx^2 + \cdots$ represents the spatial part with the positive signature.
We defined  
\begin{align}
X=x+y , \;\;\;\; \Phi = x-y,
\label{def of X and Phi in 11d}
\end{align}
for the second line of (\ref{11d metric}).
We assume that one of the light-cone directions is compactified with 
radius $R$ as in the usual context of the matrix model.

We only consider the momentum transfer in the 4-dimensional subspace 
parametrized by $(x^+, x^-, \Phi, X)$ and label the momentum as
\begin{align}
p_\mu = (p_+, p_-, p_\Phi, p_X).
\end{align}
The on-shell condition is $p_+ p_- - 2p_X p_\Phi =0$.
We also use the matrix form of the momentum,
\begin{align}
p_{\alpha \dot{\alpha}} = 
\left(
\begin{array}{cc}
p_- & \sqrt{2} p_X \\
\sqrt{2} p_\Phi & p_+ \\
\end{array}
\right),
\end{align}
where $\alpha, \dot{\alpha} =1,2$.
The on-shell condition is translated to ${\rm det}(p_{\alpha \dot{\alpha}}) =0$.
In general, $2 \times 2$ matrices with vanishing determinant can be 
expressed as 
\begin{align}
p_{\alpha \dot{\alpha}} = \lambda_\alpha \bar{\lambda}_{\dot{\alpha}},
\end{align}
in terms of two spinors $\lambda$ and $\bar{\lambda}$ called
spinor helicity variables.
For two spinors $\lambda^1_\alpha$ and $\lambda^2_\alpha$, 
we define the anti-symmetric paring by 
\begin{align}
\langle 1, 2 \rangle = \epsilon^{\alpha \beta}\lambda^1_\alpha
\lambda^2_\beta,
\end{align}
where $\epsilon^{\alpha \beta}$ is the antisymmetric tensor with 
$\epsilon^{12}=1$.

We let $p^1_\mu$ be the external momentum of
an incoming graviton and $p^2_\mu$ and $p^3_\mu$ be the momenta of
two outgoing gravitons. 
We consider the following configuration for these momenta,
\begin{align}
&p_\mu^1 = \left(0, -\frac{N_1}{R}, p^{cm}, 0 \right),
\nonumber\\
&p_\mu^2 = \left(0, \frac{N_2}{R}, p-\frac{N_2}{N_1}p^{cm}, 0 \right),
\nonumber\\
&p_\mu^3 = \left(0, \frac{N_3}{R}, -p-\frac{N_3}{N_1}p^{cm}, 0 \right),
\label{external momenta}
\end{align}
where the integers $N_i$ label the light-cone momenta and $p $ and 
$ p^{cm}$ correspond to the relative and the center-of-mass  
momentum, respectively\footnote{Though the gravitons are massless, 
we call it ``center-of-mass momentum'' in the sense that 
writing the light-cone energy as $p_+ = \frac{\vec{p}^2}{2p_-}$,  $p_-$ can be 
thought of as 
a mass of a nonrelativistic system with energy $p_+$. Thus, the ``mass '' is 
proportional to $N_i$ for each graviton. This wording is indeed 
correct when we discuss the matrix model, where the graviton with 
$p_-=N_i/R$ is described by $N_i$ coincident D0-branes, 
which have mass proportional to $N_i$.
}. For the momentum conservation, we assume $N_1=N_2+N_3$.
The corresponding spinor helicity variables are chosen as 
\begin{align}
&\lambda^1 = 
\left(
\begin{array}{c}
-\sqrt{\frac{N_1}{R}} \\
p^{cm} \sqrt{\frac{2R}{N_1}} \\
\end{array}
\right), \;\;\;\;\;\;
\bar{\lambda}^1 = 
\left(
\begin{array}{c}
\sqrt{\frac{N_1}{R}} \\
0 \\
\end{array}
\right),
\nonumber\\
&\lambda^2 = 
\left(
\begin{array}{c}
-\sqrt{\frac{N_2}{R}} \\
-\left(p-\frac{N_2}{N_1} p^{cm} \right) \sqrt{\frac{2R}{N_2}} \\
\end{array}
\right), \;\;\;\;\;\;
\bar{\lambda}^2 = 
\left(
\begin{array}{c}
-\sqrt{\frac{N_2}{R}} \\
0 \\
\end{array}
\right),
\nonumber\\
&\lambda^3 = 
\left(
\begin{array}{c}
-\sqrt{\frac{N_3}{R}} \\
\left(p+\frac{N_3}{N_1} p^{cm} \right) \sqrt{\frac{2R}{N_3}} \\
\end{array}
\right), \;\;\;\;\;\;
\bar{\lambda}^3 = 
\left(
\begin{array}{c}
-\sqrt{\frac{N_3}{R}} \\
0 \\
\end{array}
\right).
\label{choosing lambda}
\end{align}

In general, the form of three point amplitude is determined 
up to an overall constant from the 11-dimensional 
super Poincare symmetry as
\begin{align}
{\cal A}_{3} = \frac{C}{(\langle 1, 2 \rangle \langle 2, 3 \rangle 
\langle 3, 1 \rangle)^2}  
\delta(\sum_{i=1}^3 p_+^i)
\delta(\sum_{i=1}^3 p_-^i)
\delta^{(9)}(\sum_{i=1}^3 \vec{p^i})
\delta^{(16)}(\sum_{i=1}^3 \lambda_\alpha^i \bar{\eta}^i_I),
\label{3pt amplitude of gravitons}
\end{align}
where $C$ is the constant.

By changing the variables as 
\begin{align}
&\bar{\eta}^2_I = 
\sqrt{\frac{N_3}{N_1}} \bar{\eta}^r_I
-\sqrt{\frac{N_2}{N_1}} \bar{\eta}^{cm}_I,
\nonumber\\
&\bar{\eta}^3_I = 
-\sqrt{\frac{N_2}{N_1}} \bar{\eta}^r_I
-\sqrt{\frac{N_3}{N_1}} \bar{\eta}^{cm}_I,
\end{align}
and substituting our setup (\ref{choosing lambda}) 
into (\ref{3pt amplitude of gravitons}), we obtain
\begin{align}
{\cal A}_{3} = 
\frac{2C N_1^4 } {N_2^2 N_3^2} \; p^2 \;
\delta(\sum_{i=1}^3 p_+^i)
\delta(\sum_{i=1}^3 p_-^i)
\delta^{(9)}(\sum_{i=1}^3 \vec{p^i})
\delta^{(8)}(\bar{\eta}^1_I -\bar{\eta}^{cm}_I)
\delta^{(8)}(\bar{\eta}^r_I)
\label{gravity amplitude propto p2}
\end{align}
Thus, the three point amplitude is proportional to $p^2$, the square 
of the relative momentum.
In the subsequent sections, we try to reproduce this dependence 
from the matrix model side.

The amplitude with (\ref{external momenta}) preserves 1/4 supersymmetries.
The projection to this 1/4 BPS sector is given by\footnote{The condition (\ref{projection for epsilon 11})
follows from $p_\mu^i \gamma^\mu \epsilon^{(11)} =0 \; (i=1,2,3)$.}
\begin{align}
\gamma^{-} \epsilon^{(11)} =\gamma^{\Phi} \epsilon^{(11)} = 0,
\label{projection for epsilon 11}
\end{align}
where $\epsilon^{(11)}$ is a spinor parametrizing the supersymmetries.
In addition, we will consider parallel Euclidean M2-branes as source of the gravitons.
In this case, the branes break half of the 1/4 supersymmetries and 
$\epsilon^{(11)}$ satisfies
\begin{align}
\gamma^{123}\epsilon^{(11)}= -i\epsilon^{(11)},
\label{projection 2 for epsilon 11}
\end{align}
where 1, 2 and 3 are the directions to which the branes extend.

\section{BFSS matrix model}
\label{section3}

The matrix model corresponding to the signature in 
(\ref{11d metric}) is the Lorentzian model with one of bosonic matrices 
Wick-rotated to the opposite signature.
In order to define the path integral without any subtlety of convergence, 
we make a double Wick rotation for the $x^+$ and $\Phi$ directions.
Since the momentum transfer in (\ref{external momenta}) for those directions is trivial,
(\ref{gravity amplitude propto p2}) is invariant under the rotation up to an overall constant.
We thus expect that the momentum dependence of the 
partition function of the matrix model is also invariant.

We divide this double Wick rotation into two steps.
We first Wick-rotate the $x^+$ direction, which in fact is the temporal direction of the 
matrix model, and construct a supersymmetry for a localization computation in this section.
We will then make the rotation of $\Phi$ in the next section 
when we discuss quantum theory\footnote{
Without the rotation of $\Phi$, the path integral does not converge,
since the model still contains a wrong-sign bosonic matrix.
This situation is similar to the work \cite{Pestun:2007rz}}.

We use the label $M=0, 1, 2, \cdots, 9$, and $9$ is used for the 
Euclidean time and $0$ is for the wrong-sign direction.
The action is given by
\begin{align}
S =\frac{1}{g^2}\int d\tau {\rm Tr} 
\left( 
\frac{1}{2} \sum_{M=0}^8 (D_9 X_M)^2 -\frac{1}{4}\sum_{M, N=0}^8 [X_M, X_N]^2
+\frac{1}{2} \Psi^T C^\dagger \Gamma^9 D_9 \Psi
-\frac{i}{2} \sum_{M=0}^8 \Psi^T C^\dagger \Gamma^M [X_M, \Psi]
\right),
\label{on-shell action of BFSS}
\end{align}
where $D_9   = \partial_9  -i[X_9, \;\; ]$ is the covariant derivative along 
the Euclidean time with the gauge field $X_9=A$.
For simplicity, we may write $D$ instead of $D_9$ in this paper.
The model contains $N \times N$ bosonic Hermitian matrices $X_M$ and 
$N \times N$ fermionic matrices $\Psi_\alpha$, which compose a matrix-valued 
10-dimensional Majorana-Weyl spinor.
$\Gamma^M$ is the gamma matrices associated with $spin(9, 1)$.
See Appendix \ref{A Killing spinor used for localization} for the relation 
between $\Gamma^M$ and the gamma matrices in (\ref{projection for epsilon 11})
and (\ref{projection 2 for epsilon 11}).

In this section, we consider the theory 
(\ref{on-shell action of BFSS}) on a line segment $[\tau_0, \tau_1]$
and develop an off-shell supersymmetric formulation
 in the presence of boundaries. We also 
discuss boundary conditions at $\tau = \tau_0, \tau_1$, which realize
the gravitational three point amplitude.

\subsection{Supersymmetry}

On-shell supersymmetry of the BFSS matrix model is 
generated by 10-dimensional constant Majorana-Weyl spinors.
For our purpose of the localization, it is necessary to select a specific spinor $\epsilon$ that 
corresponds to $\epsilon^{(11)}$ which satisfies 
(\ref{projection for epsilon 11}) and (\ref{projection 2 for epsilon 11}).
We construct such $\epsilon$ in Appendix \ref{A Killing spinor used for localization}.
In the following discussion, we fix the parameter to this $\epsilon$.

In order to carry out the localization computation, 
we need to extend the supersymmetry to off-shell.
We first introduce 7 auxiliary fields $K^I (I = 1, 2, \cdots, 7)$, which are
$N \times N$ bosonic matrices.
We then define the off-shell supersymmetry by
\begin{align}
&\delta X^M = -\epsilon^T C^\dagger \Gamma^M \Psi,
\nonumber\\
&\delta \Psi = \frac{1}{2} F_{MN} \Gamma^{MN} \epsilon+K^I \nu_I,
\nonumber\\
&\delta K^I = (\nu^I)^T C^\dagger \Gamma^M D_M\Psi.
\label{off shell delta}
\end{align}
Here, we use the ten-dimensional notation:
\begin{align}
&F_{9M} = DX_M,
\nonumber\\
&F_{MN} = -i[X_M, X_N]  \;\;\; (M, N \neq 9),
\nonumber\\
&D_M \Psi =  -i[X_M, \Psi]  \;\;\; (M \neq 9), 
\end{align}
and $\nu_I$ are parameters determined from $\epsilon$ 
as in (\ref{def of nuI}).
If quadratic terms of $K_I$ are added to the action
(\ref{on-shell action of BFSS}), the theory 
is invariant under (\ref{off shell delta}). We will demonstrate this
invariance in the presence of boundaries in later subsections.

It is useful to change the fermionic variables by expanding 
\begin{align}
\Psi = \Psi_{M'} \Gamma^{0M'} \epsilon + \Upsilon_I \nu^I,
\end{align}
where, $M' = 1, 2, \cdots, 9$. 
The inverse transformation is given by
\begin{align}
&  \Psi_{M'} =-\frac{i}{v^0}\epsilon^T C^\dagger \Gamma_{M'} \Psi,
\nonumber\\
& \Upsilon_I
=
\frac{i}{v^0} \nu_I^T C^\dagger \Gamma^0 \Psi,
\end{align}
where $v^0:= i \epsilon^T C^\dagger \Gamma^0 \epsilon$.
In this paper, we choose the representation of $C$ so that $v^0$ is positive.
We also change the bosonic variables $(K^I, X^0)$ to
\begin{align}
&H^I  :=-iv^0(K_I -F_{9I}) - \frac{i}{2}w^{09IJK} F_{JK},
\nonumber\\
&\Phi := X^8-X^0.
\label{def of H and Phi}
\end{align}
$w^{09IJK}$ are bilinear forms made from $\epsilon$.
See Appendix \ref{Bilinear forms} for the definition of $w^{09IJK}$.

In terms of the new variables, the
supersymmetry takes a relatively simple form, 
\begin{align}
&\delta X^{M'} = -iv^0 \Psi^{M'},
\nonumber\\
&\delta \Upsilon_I = \frac{i}{v^0}H_I,
\nonumber\\
&\delta \Psi_{M'} = - D_{M'}\Phi,
\nonumber\\
&\delta H_I =  -i(v^0)^2[ \Upsilon_I, \Phi],
\nonumber\\
&\delta \Phi =0.
\end{align}
It is easy to see that $\delta^2$ gives a gauge transformation with parameter $\Phi$.

\subsection{BRST symmetry and $Q$-complex}

The gauge symmetry of the BFSS matrix model can be promoted to 
the BRST symmetry as in usual gauge theories.
By introducing the ghost fields $(C, \tilde{C}, B)$,
we define the standard BRST transformation as follows:
\begin{align}
&\delta_B X_{M'} = D_{M'} C,
\nonumber\\
&\delta_B \Upsilon_I = i\{ C, \Upsilon_I \},
\nonumber\\
&\delta_B \Psi_{M'} = i\{C, \Psi_{M'} \},
\nonumber\\
&\delta_B H_I = i[C, H_I],
\nonumber\\
&\delta_B \Phi =i[C, \Phi],
\nonumber\\
&\delta_B C = iC^2,
\nonumber\\
&\delta_B \tilde{C} = iB,
\nonumber\\
&\delta_B B=0.
\end{align}
It is easy to check that $\delta_B^2$ is vanishing as required.

We extend the supersymmetric transformation to the ghost fields by defining
\begin{align}
&\delta C= iv^0 \Phi,
\nonumber\\
&\delta \tilde{C} =\delta B=0.
\end{align}
Then, we define the combined symmetry 
\begin{align}
Q:= \delta - \delta_B
\label{def of Q}
\end{align}
 on all the fields including the ghosts. 

It is easy to see that $Q$ is nilpotent, $Q^2 =0$, and hence forms a complex.
In order to see this structure, it is useful to make redefinitions,
\begin{align}
&\tilde{\Psi}_{M'} := \Psi_{M'}-\frac{i}{v^0}D_{M'}C,
\nonumber\\
&\tilde{H}_I := H_I -v^0 \{ C, \Upsilon_I \},
\nonumber\\
&\tilde{\Phi} := \Phi-\frac{1}{v^0}C^2.
\end{align}
If we divide the fields into two groups $(Z_0, Z_1)$ and $(Z_0', Z_1')$ defined by
\begin{align}
&Z_0 := X_{M'}, \hspace{2.3cm} Z_0' := -iv^0 \tilde{\Psi}_{M'},
\nonumber\\
&Z_1:=(-iv^0 \Upsilon_I, C, \tilde{C}), \;\;\;\; Z_1':=(\tilde{H}_I, iv^0\tilde{\Phi}, -iB),
\label{def of Z}
\end{align}
we can explicitly see the $Q$-complex structure as 
\begin{align}
QZ_i =Z_i',   \;\;\;\; QZ_i' = 0.
\label{eq for Q-complex}
\end{align}

\subsection{Action with boundary terms}

Let us consider the theory on a segment $[\tau_0, \tau_1] \subset {\mathbb R}$.
The bulk action of the BFSS matrix model is defined by
\begin{align}
S =\frac{1}{g^2} \int_{\tau_0}^{\tau_1} d\tau {\rm Tr} 
\left( 
\frac{1}{4} F_{MN}F^{MN} +\frac{1}{2} \Psi^T C^\dagger \Gamma^M D_M \Psi
-\frac{1}{2} K_I K^I
\right),
\label{action of BFSS}
\end{align}
which is (\ref{on-shell action of BFSS}) plus the quadratic terms of $K_I$.
Note that the last term
has the wrong sign. So we have to integrate $K_I$ along the pure 
imaginary direction, making the analytic continuation to anti-Hermitian matrices.

If the time direction is ${\mathbb R}$ rather than the line segment, 
the action (\ref{action of BFSS}) is invariant under the off-shell supersymmetry
(\ref{off shell delta}) as well as the $Q$-symmetry (\ref{def of Q}).
However, in the presence of the boundaries, 
some surface terms do not vanish in the variation of $S$. 
In fact, the $Q$-variation of $S$ gives the following surface terms,
\begin{align}
QS =
\frac{v^0}{g^2} {\rm Tr} \left[ 
\frac{i}{2}
\left(K_I  -F_{9I}
 -\frac{1}{2v^0} w^{09IJK} F_{JK}
\right)\Psi_{I}  
-\frac{i}{2} D\Phi \Psi_{8} 
- \frac{1}{2} \Psi_{9}[\Phi, X_8]  
 - \frac{1}{2} \Upsilon_I [\Phi, X_I]
\right]^{\tau_1}_{\tau_0},
\end{align}
where the notation $[f(\tau)]^{\tau_1}_{\tau_0}=f(\tau_1)-f(\tau_0)$ is 
used. 
Thus, in general, the supersymmetry is broken at the boundaries 
for the naive bulk action (\ref{action of BFSS}).

In order to restore the supersymmetry, we introduce the following boundary terms,
\begin{align}
S_b :=  \frac{1}{g^2}{\rm Tr} 
\left[ 
-\frac{w^{09IJK}}{6v^0}X_I F_{JK}
-\frac{iv^0}{2} \Psi_9 \Psi_8 
-\frac{iv^0}{2} \Upsilon_I \Psi_I
\right]^{\tau_1}_{\tau_0}.
\label{def of boundary terms}
\end{align}
Then, one can easily show that $Q(S+S_b)=0$.
Therefore, the $Q$-symmetry 
is preserved in the theory with $S_b$.

The first term in (\ref{def of boundary terms}) is cubic in $X_I$ and 
can be regarded as the boundary Myers term \cite{Myers:1999ps}.
This term makes it possible to have an extended objects at the 
boundaries, as we will see in the next subsection.

Note also that (\ref{def of boundary terms}) 
includes $v^0$ and $w^{09IJK}$, which depend on the choice of $\epsilon$. 
Thus, for a general boundary condition, 
this formulation only respects a single supersymmetry 
generated by $\epsilon$.

However, for some special boundary conditions, 
the boundary terms have an enhanced symmetry.
For example, if we consider a boundary condition under which 
some terms of (\ref{def of boundary terms}) vanish, realizing 
$SO(3)$ symmetry which rotates $X_1, X_2$ and $X_3$, 
then the theory has 
at least 4 supersymmetries, because the commutator of the 
$SO(3)$ transformations and the 
supersymmetry of $\epsilon$ yields other independent 
supersymmetries. 
These enhanced supersymmetries are those satisfying
the first 3 conditions of (\ref{five conditions for epsilon}).
The boundary condition we will discuss in section 
\ref{Boundary condition}
are precisely for this case.

In fixing the gauge, we also add a $\delta_B$-exact terms to the action as 
usual. We will apply this procedure when we discuss the localization in the next section.

\subsection{BPS solution}

The bulk equation of motion has a special kind of solution which 
preserves the supersymmetry $\delta$.
This is the solution to the Nahm equation,
\begin{align}
DX_a + \frac{i}{2} \epsilon_{abc} [X_b, X_c] =0.
\label{Nahm eqation with Xa}
\end{align}
The equation is formally solved in $X_9=A=0$ gauge by 
\begin{align}
X_a = \frac{1}{\tau_* -\tau} J_a,
\label{Nahm solution}
\end{align}
where $\tau_*$ is a real constant and $J_a$ are the $N$-dimensional 
representation matrices 
of $SU(2)$ generators satisfying $[J_a, J_b] = i\epsilon_{abc}J_c$.

The bulk action is divergent for 
the solution (\ref{Nahm solution}) in general. However, if $\tau_*$ is 
at the boundary, the divergence is cancelled by the boundary 
Myers term in (\ref{def of boundary terms}). 
We will demonstrate this in the localization in the next section.

There is another class of BPS solution obtained by solving 
$DX_a - \frac{i}{2} \epsilon_{abc} [X_b, X_c] =0$. The change of the sign of
the second term
is crucial in our context because solutions of this equation
do not preserve the supersymmetry of 
$\epsilon$. While (\ref{Nahm eqation with Xa}) describes a decay of branes, 
the one with the minus sign describes a merge of branes.
Thus, with our choice of the supersymmetry, the ``orientation'' is introduced 
in our formulation.

The solution (\ref{Nahm solution}) 
describes Euclidean M2-branes at $\tau = \tau_*$ extending to the $1,2$ and $3$ directions  
\cite{Diaconescu:1996rk}.
By imposing the boundary condition such that 
the fields behave as (\ref{Nahm solution})  near the two 
edges of the line segment $[\tau_0, \tau_1]$ and taking $\tau_0 \rightarrow -\infty$
and $\tau_1 \rightarrow \infty$, we can realize asymptotic states with gravitational 
sources given by the M2-branes.
We discuss the boundary condition more precisely below.

\subsection{Boundary condition for $SU(N)$ modes}
\label{Boundary condition}

The gauge group we consider is $U(N)$.
Since the overall $U(1)$ modes are free fields
decoupled from the other $SU(N)$ modes, 
they can be integrated out without using the localization.
See appendix \ref{Path integral of U1 modes}.
Below, we only focus on the $SU(N)$ modes and 
specify boundary conditions relevant to the gravity amplitudes.

For the problem of the three point amplitude, we consider a single M2-brane at $\tau_0$ 
and two M2-branes at $\tau_1$. 
This corresponds to the case where
the representation of $J_a$ is a single irreducible representation at $\tau_0$ and 
is the direct sum of two irreducible representations at $\tau_1$.
Thus, the relevant boundary condition is
\begin{align}
X_a(\tau) \sim 
\begin{cases}
\frac{1}{\tau_0 -\tau} J_a^{[N_1]}   \;\;\; & (\tau \rightarrow \tau_0) \\
\frac{1}{\tau_1-\tau } J_a^{[N_2]}\oplus J_a^{[N_3]} \;\;\; & (\tau \rightarrow \tau_1) \\
\end{cases},
\label{bc for Xa}
\end{align}
where  $J_a^{[n]}$ represent the generators in the $n$-dimensional irreducible representation.
$N_1=N$ is the total matrix size and $N_2, N_3$ are positive 
integers satisfying $N_2 + N_3 =N$ and labeling the 
light-cone momenta of the two objects at $\tau_1$.

We define
\begin{align}
X := X^8 + X^0.
\end{align}
The matrices $X$ and $\Phi$ in (\ref{def of H and Phi})
describe the directions in (\ref{def of X and Phi in 11d}) in 11 dimensions.
We impose the boundary conditions
\begin{align}
&D\Phi \rightarrow 0  \;\;\;\; (\tau \rightarrow \tau_{0}, \tau_1),
\nonumber\\
&DX \rightarrow 
\begin{cases}
 0   & (\tau \rightarrow \tau_0) \\
 \frac{p_2}{N_2} {\bf 1}_{N_2} \oplus \frac{p_3}{N_3} {\bf 1}_{N_3} & (\tau \rightarrow \tau_1) \\
\end{cases},
\label{bc for X Phi}
\end{align}
with $p_2, p_3$ real numbers. In order for the $DX$ to be traceless, we have
$p_2+p_3 =0$. 
Note that $DX$ and $p:=p_2-p_3$ correspond to $p_{\Phi}$ and 
the relative momentum in 11 dimensions, respectively.

There are other bosonic fields, $X_4, X_5, X_6, X_7$ and $H_I$.
Since we are not interested in the motion along the 4,5,6,7 directions, 
we fix them as 
\begin{align}
X_I= 0  \;\;\;\; (I=4,5,6,7),
\label{bc for XI}
\end{align}
at the both boundaries. 
For the auxiliary fields $H_I$, we set 
\begin{align}
D\tilde{H}_I =0  \;\;\;\; (I=1, 2, \cdots, 7 ).
\label{bc for HI}
\end{align}  

In order for the $Q$-complex structure to be well-defined,
we also require that fields in the same doublet of $Q$-symmetry should 
obey the same kind of boundary conditions. This leads to the conditions
\begin{align}
&D\tilde{\Psi}_a +i[X_a, \tilde{\Psi}_9]+ i\epsilon_{abc}[X_a, \tilde{\Psi}_c] =0,
\nonumber\\
&\tilde{\Psi}_I  =0 \;\;\;\; (I=4,5,6,7),
\nonumber\\
&D\tilde{\Psi}_8 =0,
\nonumber\\
&DC =0,
\nonumber\\
&D\Upsilon_I =0 \;\;\;\; (I=1,2, \cdots, 7).
\label{bc for fermions}
\end{align}
The first condition is obtained as the $Q$-transform of
 (\ref{bc for Xa}). $\tilde{\Psi}_a$ is the supersymmetric partner 
of the fluctuation of $X_a$ around the Nahm pole, which is finite 
at the pole,  so we also assume that $\tilde{\Psi}_a$ are finite 
at the Nahm pole.
We make free the boundary conditions for the other fields, 
$X_9 (=A), B, \Psi_9$ and $\tilde{C}$.

Although we are interested in the theory with a general matrix size, 
because of some technical difficulty in the localization, 
we will restrict ourselves to the case of $N=2$ and $\tau_1 \rightarrow \infty$
in the next section. 
In this case, the Euclidean time direction is the half line $[\tau_0, \infty )$
and the only possible choice is $N_2 = N_3=1$. 
The condition (\ref{bc for Xa}) becomes
\begin{align}
X_a(\tau) \sim 
\begin{cases}
\frac{1}{\tau_0 -\tau} J_a^{[2]}   \;\;\; & (\tau \rightarrow \tau_0) \\
0 \;\;\; & (\tau \rightarrow \infty) \\
\end{cases}.
\label{bc for X with N=2}
\end{align}
The branes at $\tau_1$ are just two gravitons (D0-branes) sitting at the origin of the 
space of $X_{1, 2, 3}$.
Then, the two branes at $\tau_1$ are only separated by 
the condition of $X$ in (\ref{bc for X Phi}).
Actually, we can relax the boundary condition at $\tau_1$ 
such that $X_{1}$ and $X_2$ go to zero but 
\begin{align}
X_{3} \rightarrow - |c| J_3^{[2]}  \;\;\;\; (\tau \rightarrow \infty)
\label{bc for X3}
\end{align}
with $c$ a nonzero real constant. Then, the two gravitons are 
separated by the distance $|c|$ in the $X_3$ direction.
The constant $c$ plays the role of some regularization parameter in the 
localization.
For a comparison with the amplitude on the gravity side,
one should take $|c|\ll g^{2/3}$
because the scattering amplitude \eqref{gravity amplitude propto p2} 
does not involve the 3-direction.

In summary, the boundary condition we consider is 
(\ref{bc for Xa}), (\ref{bc for X Phi}), (\ref{bc for XI}), (\ref{bc for HI}), 
and (\ref{bc for fermions}). 
For $N=2$ and $\tau_1 = \infty$, we replace (\ref{bc for Xa}) with 
 (\ref{bc for X with N=2}) modified by (\ref{bc for X3}).
We will see that there exists an interesting 
localization saddle satisfying these conditions.

\section{Localization}
\label{section4}

In this section, we consider the $U(2)$ BFSS matrix model on $[\tau_0, \infty)$
and perform the localization with the boundary condition 
discussed in section \ref{Boundary condition}.
Since the overall $U(1)$ modes are free particles and give trivial momentum conservation 
of the center-of-mass mode (see appendix \ref{Path integral of U1 modes}), 
here we only consider the $SU(2)$ modes.

\subsection{Potential for localization}


We introduce the functional,
\begin{align}
V&=\int d\tau {\rm Tr} \left[ \overline{\delta \Psi}^T \Psi \right] =\int d\tau
{\rm Tr} \left[ 
-\frac{1}{2} F_{MN}
\epsilon^T iC^\dagger \Gamma^0
\Gamma^{MN}\Psi  -K^{I} \nu_I^T iC^\dagger \Gamma^0\Psi
\right],
\end{align}
and the functional at the boundary,
\begin{align}
V_{b} = i {\rm Tr}\left[ (D X) C \right]^{\tau_1}_{\tau_0}.
\end{align}
For gauge fixing, we also introduce 
\begin{align}
V_{gh} = \int d\tau {\rm Tr} \left( \tilde{C}(\partial X_9 -i[\hat{X}_a, X_a]) + \frac{i\xi}{2}\tilde{C}B  \right),
\end{align}
where $\hat{X}_a$ is some fixed configuration and $\xi$ is a positive parameter. 
We choose $\hat{X}_a$ to be the saddle point configuration discussed in 
the next subsection. Then, the above $V_{gh}$ gives the BRST formalism for the  
background gauge, where the background is given by the saddle point configuration.

In the localization, we add $tQ(V+V_b+V_{gh})$ to the action. 
From the general argument of localization, the partition function does not 
depend on $t$ and one can take $t \rightarrow \infty$ in evaluating the 
partition function.
Then, the path integral is dominated by the saddle point configurations of
 $Q(V+V_b+V_{gh})$. 
Schematically, the partition function is reduced to the form,
\begin{align}
Z = \sum_{i } Z^{(i)}_{\rm 1-loop} e^{-S_{\rm cl}^{(i)}},
\end{align}
where $i$ labels saddle points and 
$S_{\rm cl}^{(i)}$ and $Z^{(i)}_{\rm 1-loop}$ are the 
bosonic part of the classical action and 
the 1-loop determinant at the $i$th saddle point, respectively.

Let us compute the 
explicit form of  $Q(V+V_b+V_{gh})$.
Since $V_{gh}$ only gives the usual gauge fixing, 
we compute $Q(V+V_b)$ below.
After some calculations, we can rewrite $V$ as
\begin{align}
V&= \int d\tau {\rm Tr} \left[
v^0 (F_{8N'}-F_{0N'}) \Psi^{N'}
-\left(v^0 K^I +v^0F_{9I}- \frac{1}{2}w^{09IJK}F_{JK}
\right)
\Upsilon_I
\right],
\end{align}
where $M', N' = 1, 2, \cdots, 9$.
We find that the bosonic part of $Q(V+V_b)$ is 
\begin{align}
Q (V+V_{b})|_{\rm boson}
=
\int d\tau {\rm Tr}
&\left[
-v^0 (D^{N'}D_{N'}X) \Phi - v^0 K_I K^I 
+\frac{1}{v^0} \left(v^0F_{9I}- \frac{1}{2}w^{09IJK}F_{JK}\right)^2
\right].
\label{delta V+ Vb}
\end{align}
Note that $K_I$ has the wrong sign and we integrate $K_I$ along the imaginary axis.
Then, the saddle point of $K_I$ is simply $K_I =0$.
In the following, we will investigate saddle-point configurations of the other fields.

\subsection{Saddle point}

While the last term in (\ref{delta V+ Vb}) is positive definite, the first term is not.
So the naive integration of $\Phi$ gives divergence. 
In order to avoid the divergence, we make the analytic continuation 
of $\Phi$ and integrate it over the imaginary axis
as we did for $K_I$. This is possible since the boundary 
condition of $\Phi$ is the simple Neumann  condition (\ref{bc for X Phi}).
Then, the term with $\Phi$ in (\ref{delta V+ Vb})
becomes pure imaginary, which in turn gives a phase factor in 
the path integral.
Since the phase is proportional to $t$,
in the large-$t$ limit, only paths with a constant phase survive. 
Thus, the saddle point is given by the solution to the equations
\begin{align}
&D^{N'}D_{N'}X= D^{N'}D_{N'}\Phi'=0,
\nonumber\\
&v^0F_{9I}- \frac{1}{2}w^{09IJK}F_{JK} = 0.
\label{saddle eq}
\end{align}
Here, $\Phi'$ is the analytic continuation of $\Phi$. Note that
the phase is not only stationary but vanishing on the saddle.

The functional $V$ is well-defined for general matrix size $N$ 
and for general $\tau_0, \tau_1$. 
Then, the saddle point equations (\ref{saddle eq}) are 
also valid for general $N$.
However, the second equation of (\ref{saddle eq}) involves the Nahm 
equation and it is known to be a very difficult problem to solve this equation
for general $N$.
Thus, we restrict ourselves to the case with $N=2$ and $\tau_1= \infty$ below, 
in which we can solve (\ref{saddle eq}) \cite{Braden:2022ljy}.

For $N=2$ and $\tau_1= \infty$,
we solve these equations in appendix \ref{Solving saddle point equations}.
We find that the unique solution to the equations (\ref{saddle eq})
with the boundary conditions discussed in section \ref{Boundary condition}
is given by
\begin{align}
&\hat{X}_{a} = \frac{c}{\sinh(c(\tau_0-\tau))} J^{[2]}_{a}  \;\;\; (a=1, 2),
\nonumber\\
&\hat{X}_3 =  \frac{c}{\tanh(c(\tau_0-\tau))} J_3^{[2]},
\nonumber\\
&\hat{X}_I = 0 \;\;\; (I=4,5,6,7),
\nonumber\\
&\hat{X}= -\frac{p}{|c|} \left( 
1 - \frac{c(\tau_0-\tau)}{\tanh(c(\tau_0-\tau))} 
\right) J_3^{[2]},
\nonumber\\
&\hat{\Phi}' = 0,
\label{solution for N=2}
\end{align}
in the temporal gauge $X_9=A=0$, 
where $p =p_2-p_3$ is the relative momentum of two D-particles at 
$\tau = \infty$. 
Note that the above solution exponentially changes its behavior 
around $\tau = \tau_0 + \frac{1}{|c|}$. Thus, $c$ can be regarded 
as the time at which a single brane decays into two D-particles.

\subsection{Evaluation of classical action}
Let us compute the bosonic part of the classical action $S$ at the saddle point.

Since $X$ is always accompanied by $\Phi'$ in the action, and $\Phi'$ 
is zero at the saddle point (\ref{solution for N=2}), 
$X$ and $\Phi'$ do not contribute in this calculation.
$X_{4, 5, 6, 7}$ are also vanishing at the saddle point. 
So the only contribution comes from $X_{1, 2, 3}$.

Since the configuration (\ref{solution for N=2}) for $X_a$ 
diverges at $\tau =\tau_0$, 
we introduce a cutoff making the integration range $[\tau_0 + \kappa, \infty)$ with 
small $\kappa >0$.
Then, we can evaluate the kinetic terms as
\begin{align}
\int_{\tau_0+ \kappa}^{\infty} d\tau \frac{1}{2}{\rm Tr} (\partial \hat{X}_a)^2
=
\int_{\kappa}^{\infty} d\tau \frac{1}{4}
\left( 
 \frac{2c^4 \cosh^2(c\tau) }{ \sinh^4(c\tau)}
+ \frac{c^4}{ \sinh^4 (c\tau)}
\right),
\end{align}
and the potential terms as
\begin{align}
-\int_{\tau_0+ \kappa}^{\infty} d\tau \frac{1}{4}{\rm Tr} [\hat{X}_a, \hat{X}_b]^2
&=
\int_{\kappa}^{\infty} d\tau \frac{1}{4} 
\left( 
2\frac{c^2}{\sinh^2(c\tau)} \frac{c^2}{\tanh^2(c\tau)}
+\frac{c^4}{\sinh^4(c\tau)}
\right).
\end{align}
Thus, the total classical action is
\begin{align}
S &= \frac{c^4}{2g^2}
\int_{\kappa}^{\infty} d\tau 
 \frac{2 \cosh^2(c\tau)+1 }{ \sinh^4(c\tau)}
= 
\frac{c^3}{2g^2} \frac{\cosh(c \kappa)}{ \sinh^3(c\kappa)}.
\label{saddle bulk action}
\end{align}

We then evaluate the boundary term (\ref{def of boundary terms}).
The only contribution comes from the first term of 
(\ref{def of boundary terms}) at $\tau = \tau_0$, which
is again replaced by $\tau= \tau_0 + \kappa$.
Then, we find that
\begin{align}
S_b = 
-\frac{c^3}{2g^2} \frac{\cosh(c \kappa)}{ \sinh^3(c\kappa)}.
\label{saddle boundary action}
\end{align}

By adding (\ref{saddle boundary action}) to the bulk action,
(\ref{saddle bulk action})
we find that the total action is just vanishing,
\begin{align}
S+S_b =0,
\end{align}
at the saddle point.

\subsection{Evaluation of 1-loop determinant}

We expand $V+V_b +V_{gh}$ around the saddle point and 
denote by $V^{(2)}$ its quadratic part of fluctuation.
Schematically, $V^{(2)}$ takes the following form.
\begin{align}
V^{(2)} =(Z_0', Z_1) 
\left(
\begin{array}{cc}
D_{00} & D_{01} \\
D_{10} & D_{11} \\
\end{array}
\right)
\left(
\begin{array}{c}
Z_0 \\
Z_1' \\
\end{array}
\right),
\label{schematic form of V2}
\end{align}
where $Z_i$ and $Z'_i$ are the variables defined in (\ref{def of Z}) and 
$D_{ij}$ are some linear differential operators.
Then, by acting $Q$, we obtain
\begin{align}
QV^{(2)} 
&=
(0, Z'_1) 
\left(
\begin{array}{cc}
D_{00} & D_{01} \\
D_{10} & D_{11} \\
\end{array}
\right)
\left(
\begin{array}{c}
Z_0 \\
Z_1' \\
\end{array}
\right)
-
(Z_0', Z_1) 
\left(
\begin{array}{cc}
D_{00} & D_{01} \\
D_{10} & D_{11} \\
\end{array}
\right)
\left(
\begin{array}{c}
Z_0' \\
0 \\
\end{array}
\right)
\nonumber\\
&=
Z_1'(D_{10}Z_0 +D_{11}Z_1') -(Z_0'D_{00} +Z_1D_{10})Z_0'
\label{QV quadratic}
\end{align}
Note that $Z_0$ and $Z_1$ are linear in this expression. 
Thus, by integrating them with suitable Wick rotations, 
they produce delta functions constraining $Z_0'$ and $Z_1'$.
Then, the 1-loop determinant coming from $Z_0'$ and $Z_1'$ takes the form,
\begin{align}
\left( \frac{{\rm det}D_{00}|_{{\rm Ker}D_{10}} }{{\rm det}D_{11}|_{{\rm Coker}D_{10}}}
\right)^{\frac{1}{2}},
\label{1-loop det}
\end{align}
up to an overall constant.
See appendix \ref{Derivation of 1-loop det} for a detailed derivation of this expression.
In the following, we will compute (\ref{1-loop det}) and show that it is proportional to 
$p^2$, where $p$ is the relative momentum of two D-particles at $\tau = \infty$.

We first consider the numerator of (\ref{1-loop det}).
For $\tilde{\Psi}_{M'} \in {\rm Ker}D_{10}$, each coefficient of $C$, $\tilde{C}$ and
$\Upsilon_I$ in $Z_1 D_{10} Z_0'$ is vanishing, where $Z_1 D_{10} Z_0'$ is explicitly given by
\begin{align}
Z_1 D_{10} Z_0' =
\int^\infty_{\tau_0} d\tau \; iv^0 {\rm Tr} &
\left[
-2C ([\hat{X}_8, \partial \tilde{\Psi}_9] +2[\partial \hat{X}_8, \tilde{\Psi}_9] 
-i \partial^2 \tilde{\Psi}_8)
\right.
\nonumber\\
&+2i C ([\hat{X_8}, [\hat{X}^{N'}, \tilde{\Psi}_{N'}] ] +2[[\hat{X}^{N'}, \hat{X}_8], \tilde{\Psi}_{N'}]
- [\hat{X}^{N'}, [\hat{X}_{N'}, \tilde{\Psi}_8]])
\nonumber\\
&\left.
-2v^0 \Upsilon_I \left( i[\hat{X}_I, \tilde{\Psi}_9] + \partial \tilde{\Psi}_I + 
i \frac{w^{09I a J}}{v^0} [\hat{X}_a, \tilde{\Psi}_J] \right)
+ \tilde{C}(\partial \tilde{\Psi}_9 -i[\hat{X}_a, \tilde{\Psi}_a])
\right].
\label{Z1D10Z0'}
\end{align}
By solving these 
conditions (see appendix \ref{computation of ker}), 
we find that ${\rm Ker}D_{10}$ consists of the configuration (\ref{zero mode fermions}). 
Thus, ${\rm Ker}D_{10}$ is generated by four real elements, which are in 
the subspace spanned by $\tilde{\Psi}_{a}$ and $\tilde{\Psi}_9$.
In this subspace, the operator $D_{00}$ acts as 
\begin{align}
D_{00}
\left(
\begin{array}{cc}
\tilde{\Psi}_a  \\
\tilde{\Psi}_9  \\
\end{array}
\right)
\propto
[\hat{X},   
\left(
\begin{array}{cc}
\tilde{\Psi}_a  \\
\tilde{\Psi}_9  \\
\end{array}
\right)
],
\label{[X, Psi]}
\end{align}
which we find from the following explicit form,
\begin{align}
Z_0' D_{00}Z'_0
= \int^\infty_{\tau_0} d\tau \; i(v^0)^2 {\rm Tr}
\left[ 
2(\partial \tilde{\Psi}_8 +i[\hat{X}_8, \tilde{\Psi}_9]) \tilde{\Psi}_9
- 2i\sum_{N= 1}^8 ([\hat{X}_N, \tilde{\Psi}_8] -[\hat{X}_8, \tilde{\Psi}_N])
\tilde{\Psi}_N
\right].
\label{Z0'D00Z0'}
\end{align}
Since $\hat{X}$ is proportional to the relative momentum $p$, 
the functional determinant of $D_{00}$ restricted to
${\rm Ker}D_{10}$ is proportional to $p^{{\rm dim}({\rm Ker}D_{10})/2}=p^2$. 
Hence, we find that the numerator of (\ref{1-loop det}) is proportional to
$p^2$.

In appendix \ref{computation of coker D10}, we also compute
${\rm Coker}D_{10}$ and find that ${\rm Coker}D_{10}=\{0\}$.
Therefore, we find that the $p$-dependence of the 
1-loop determinant is finally given by
\begin{align}
Z_{\rm 1-loop} \propto p^2.
\end{align}
This exactly agrees with the $p$-dependence of the gravity three-point 
amplitude (\ref{gravity amplitude propto p2}).

\section{Summary and discussion}
\label{section5}

In this paper, we considered the graviton three-point amplitude 
(\ref{gravity amplitude propto p2}) in M-theory
focusing especially on the dependence on the relative momentum $p$ 
of two outgoing gravitons.
By applying the localization to $U(2)$ BFSS matrix model with 
a proper boundary condition, which corresponds 
to the three point amplitude 
on the gravity side, we showed that the partition function of the matrix model has 
exactly the same $p$-dependence.
This result gives strong evidence of the conjecture that the 
matrix model is a nonperturbative formulation of M-theory.

In our computation, we first established 
the formulation of the BFSS matrix model on a finite line segment, 
by implementing the appropriate boundary terms 
(\ref{def of boundary terms}).
This formulation keeps at least one exact off-shell supersymmetry 
for arbitrary boundary conditions. 
The number of supersymmetries can be 
 enhanced depending on the boundary condition, and the theory preserves 4 supersymmetries for our choice 
of the boundary condition.

We then applied the localization by using the off-shell supersymmetry.
Because of some technical difficulty which we explain shortly, 
we restricted ourselves to the case with $N=2$ on the half line 
$[\tau_0, \infty )$ by taking one of the edges to infinity.  
In this case, we succeeded in performing the localization 
and found the agreement of the $p$-dependence.

Several caveats are in order when comparing with the amplitude.
First, our result is obtained at small $N$, and 
therefore corresponds to the DLCQ of M-theory 
with fixed total light-cone momentum.
In general, amplitudes in the DLCQ formalism are different from the 
flat space amplitudes. However, we assumed that the three-point 
amplitude is an exception, because the result in
\cite{Herderschee:2023pza} suggests that the amplitude takes 
the same form even for finite $N$.
Since our result is exact and valid even in the strong coupling limit, where
the compactification radius becomes infinity,
we can then compare our result with the flat space amplitude.
Note that in the strong coupling limit with fixed $N$, 
the light-cone momenta approach zero. 
However, this only scales the overall factor of the three-point amplitude 
and the $p$-dependence should remain unchanged.


Second, we discuss the correspondence to the work 
\cite{Herderschee:2023pza}. 
Because our setup including boundary conditions differs
from the one in \cite{Herderschee:2023pza},  
a precise comparison is not possible; nevertheless, there are both similarities and differences.
In \cite{Herderschee:2023pza}, it is assumed that 
the $\tau$ direction can be separated into three parts with 
$\tau \ll \tau_*, \tau \sim \tau_*$ and $\tau \gg \tau_*$
with some $\tau_* \in {\mathbb R}$, such that
nontrivial interaction occurs only in the middle part and 
the theory behaves as free theory in the other two parts.
Our result shows a very similar behavior.
In general, a solution to the Nahm equation 
exponentially changes its behavior \cite{Braden:2022ljy} and thus, 
the region over which the solution undergoes significant variation 
is in a finite interval.
Indeed, in our case with $N=2$, 
the only contributing modes to the 1-loop determinant, 
which are given by \eqref{zero mode fermions}, 
vanish at $\tau\gg \tau_*\sim \tau_0+1/|c|$ and also at $\tau\sim\tau_0$
when they are acted on by the operator $D_{00}$ as in \eqref{[X, Psi]}.
Therefore, since we consider $1/|c|\gg g^{-2/3}$ for the comparison with the amplitude,
the contribution to the partition function comes only from the intermediate region, 
which is identified as the amplitude in \cite{Herderschee:2023pza}.
We here remark that this local structure of the partition function 
partially justifies our comparison with the amplitude on the flat space.

In \cite{Herderschee:2023pza}, 
$\tau_*$, which parametrizes the location of 
the intermediate region, is integrated over the line segment.
On the other hand, in our computation, this position is 
given by $\tau_0+1/|c|$ and is completely fixed by the 
boundary condition. This is a difference of our work and 
\cite{Herderschee:2023pza}. 
However, this is likely a peculiarity of the $N=2$ case.
For general $N$, solutions of the Nahm equation have nontrivial moduli 
in general and integration over the moduli similar to that in \cite{Herderschee:2023pza}
will arise in our framework as well.
Investigating this correspondence at larger $N$ 
remains an important problem.

While most part of our computation 
including the supersymmetric formulation mentioned above
 works for arbitrary $N$, 
there is a difficulty in finding the saddle point configuration for general $N$.
Our result shows that the saddle point equation is reduced to the Nahm
equation on a line segment (or a half line). 
In order to sum up all saddle point configurations
for localization, we have to find all solutions to the Nahm
equation for given boundary conditions.
While this is possible for $N=2$ with our setup, 
no general method is known for general $N$.
Once this point is overcome, it should be possible to extend our computation 
to general $N$.

There are several earlier studies 
\cite{Braden:2022ljy, Bachas:2000dx,Yee:2003ge,Kovacs:2015xha}
on the Nahm equation and 
one may be able to use those techniques for the localization. 
It is a challenging problem to reproduce not only the $p$-dependence
but also the $N_i$-dependence of the amplitude (\ref{gravity amplitude propto p2})
from the exact computation of the matrix model.
Our results are also applicable to higher point amplitudes.
It is interesting future work to investigate whether the general 
amplitudes are reproduced from the matrix model.
Even at finite $N$, it might be possible to compare the matrix model and 
the non-Lorentzian gravity \cite{Bergshoeff:2018yvt,Maskalaniec:2026vlk}.

In this paper we have focused on flat space, 
but there is also a matrix model for the pp-wave background \cite{Berenstein:2002jq}.
In that case the relevant matrix model is massive, which makes 
localization computation more tractable 
\cite{Asano:2012zt, Asano:2017nxw}\footnote{See 
\cite{Asano:2017nxw,Asano:2014vba,Asano:2014eca,Asano:2017xiy}
 for applications of this computation.
See also \cite{Hartnoll:2024csr,Komatsu:2024ydh,Hartnoll:2025ecj}
 for the zero-dimensional counterpart.}. It would therefore be interesting to investigate whether a similar correspondence of amplitudes can be observed in the BMN matrix model using localization.

\section*{Acknowledgments}

We thank Ruka Kato for discussions at early stage of this work.
We thank Hidehiko Shimada for valuable discussions. 
The work of Y.~A.~and G.~I.~was supported by
 JSPS KAKENHI Grant Number JP24K07036 and JP23K03405, 
respectively.
The work of Y.~M.~was supported by JST SPRING Grant Number JPMJSP2124.

\begin{appendix}
\numberwithin{equation}{section}
\setcounter{equation}{0}

\section{Killing spinors}

In this appendix, we introduce a Killing spinor, which we use for the localization.
We also introduce Killing spinors for the off-shell supersymmetry.

Let $\Gamma^M (M=0, 1,2, \cdots, 9)$ be the 10-dimensional 
($32 \times 32$) gamma matrices satisfying 
$\{\Gamma^M, \Gamma^N\} = \eta^{MN}{\bf 1}_{32}$.
We denote by $C$ the charge conjugation matrix satisfying
$C^\dagger \Gamma^M  = -(\Gamma^M)^T C^\dagger$ and
$C^T =-C$.
We also define the chirality matrix 
\begin{align}
\Gamma^{11} = -\Gamma^{012 \cdots 9}.
\end{align}

$\Gamma^M$ is related to the gamma matrices in (\ref{projection for epsilon 11})
and (\ref{projection 2 for epsilon 11}) in 11 dimensions as
\begin{align}
&\gamma^I = -i \Gamma^{9I}  \;\;\; (I=1,2, \cdots, 7), 
\nonumber\\
&\gamma^\pm = 
-\frac{i}{\sqrt{2}} \Gamma^9  ({\bf 1}_{32} \pm \Gamma^{11}),
\nonumber\\
&\gamma^\Phi = -i \Gamma^{98}({\bf 1}_{32}-\Gamma^{80}),
\nonumber\\
&\gamma^X = -i \Gamma^{98}({\bf 1}_{32}+\Gamma^{80}).
\end{align}

\subsection{A Killing spinor used for localization}
\label{A Killing spinor used for localization}

We focus on the ten-dimensional constant spinor with 32 components
which satisfies the following relations:
\begin{align}
&\Gamma^{11} \epsilon = \epsilon,
\nonumber\\
&\Gamma^{9123} \epsilon = \epsilon,
\nonumber\\
&\Gamma^{80} \epsilon =\epsilon,
\nonumber\\
&\Gamma^{1245} \epsilon = \epsilon,
\nonumber\\
&\Gamma^{1346} \epsilon = \epsilon.
\label{five conditions for epsilon}
\end{align}
Here, we let $\epsilon$ be Grassmann even.
There is a unique solution to the equations (\ref{five conditions for epsilon})
(up to the normalization factor),
since the five matrices on the left-hand sides 
 mutually commute and each of them gives an independent grading $\pm 1$
and halves the number of independent components of $\epsilon$.
The first three conditions in (\ref{five conditions for epsilon}) correspond
to the conditions imposed on the gravity side,
(\ref{projection for epsilon 11}) and (\ref{projection 2 for epsilon 11}).
The fourth and fifth conditions are our convention to single out just one 
component out of 32.

The spinor $\epsilon$ satisfying the above five conditions also satisfies
the following equations:
\begin{align}
\epsilon&= \Gamma^{3459} \epsilon 
=\Gamma^{2356} \epsilon 
=\Gamma^{9246} \epsilon
=\Gamma^{1569} \epsilon
=\Gamma^{4567} \epsilon 
=\Gamma^{9367} \epsilon 
\nonumber\\
&=\Gamma^{9257} \epsilon 
=\Gamma^{1276} \epsilon
=\Gamma^{9147} \epsilon 
=\Gamma^{1357} \epsilon 
=\Gamma^{7234} \epsilon.
\label{additional equalities for epsilon}
\end{align}

\subsection{Bilinear forms}
\label{Bilinear forms}

For $\epsilon $ satisfying (\ref{five conditions for epsilon}),
we define the bilinear forms,
\begin{align}
v^M := \epsilon^{T} iC^\dagger \Gamma^M \epsilon.
\end{align}
It turns out that the only nonvanishing components of $v^m$ are $v^0$ and $v^8$
and they are equal to each other,
\begin{align}
v^0 = v^8,   \;\;\;\;\; v^M=0 \;(M\neq 0, 8).
\end{align}
This is proved as follows. Since $C^\dagger \Gamma^{MNP}$ is anti-symmetric, 
we have the identity,
\begin{align}
\epsilon^{T} iC^\dagger \Gamma^{MNP} \epsilon= 0.
\label{triple gamma is zero}
\end{align}
By using this identity, we have for example,
\begin{align}
v^1 &=\epsilon^{T} iC^\dagger \Gamma^{1} \epsilon
=\epsilon^{T} iC^\dagger \Gamma^{1} \Gamma^{9123}\epsilon
=-\epsilon^{T} iC^\dagger \Gamma^{923}\epsilon =0.
\end{align}
We can show $v^M=0 (M\neq 0, 8)$ in the same way.
Similarly for $v^0$, we have
\begin{align}
v^0 = \epsilon^{T} iC^\dagger \Gamma^0 \epsilon =
\epsilon^{T} iC^\dagger \Gamma^0 \Gamma^{80} \epsilon
=\epsilon^{T} iC^\dagger \Gamma^{8} \epsilon =v^8.
\end{align}
$v^0$ and $v^8$ are vanishing if and only if $\epsilon =0$.

Next, we define another set of bilinear forms,
\begin{align}
w^{0MNPQ} :=\epsilon^{T} iC^\dagger \Gamma^{0MNPQ} \epsilon.
\end{align}
The nonvanishing components of $w^{0MNPQ}$ are the following:
\begin{align}
v_0&=w^{09123}=
w^{09147} =
-w^{09156} =
w^{09246} = 
w^{09257} = 
-w^{09345} =
w^{09367}  
\nonumber\\
&=w^{01245} =
-w^{01267} =
-w^{02347} =
w^{02356} =
-w^{03146} =
-w^{03157} = 
w^{04567}.
\label{values of w}
\end{align}
Let us illustrate the derivation of (\ref{values of w}).
By using the third equation in (\ref{five conditions for epsilon}), 
we have $w^{08NPQ}=-\epsilon^{T} iC^\dagger \Gamma^{NPQ} \epsilon$
but this is zero because of (\ref{triple gamma is zero}).
Thus, it suffices to consider $M, N, P, Q =1, 2, \cdots, 7, 9$.
For $(M, N,P,Q)=(9,1,2,3)$ for example, we have
\begin{align}
w^{09123} =\epsilon^{T} iC^\dagger \Gamma^{09123} \epsilon = 
\epsilon^{T} iC^\dagger \Gamma^{0} \epsilon = v^0.
\end{align}
(\ref{values of w}) can be obtained by exhaustively computing all the cases of indices
with the relations (\ref{additional equalities for epsilon}).

\subsection{Off-shell Killing spinors}
\label{Off-shell Killing spinors}

We introduce 
additional parameters $\nu^I (I=1,2, \cdots, 7)$ to extend the supersymmetry to off-shell.
They have to satisfy the following 
relations to realizes the closure of off-shell supersymmetry:  
\begin{align}
&\epsilon^T C^\dagger \Gamma^M \nu^ I =0,
\nonumber\\
&\nu_I^T C^\dagger \Gamma^M \nu^J = \epsilon^T C^\dagger \Gamma^M \epsilon \delta_I^J,
\nonumber\\
&\frac{1}{2}(\epsilon^T C^\dagger \Gamma_M \epsilon )
\left( \frac{1+\Gamma^{11}}{2}\Gamma^M C \right)_{\alpha \beta} = \epsilon_\alpha \epsilon_\beta
+\nu_\alpha^I \nu_{\beta I}.
\label{condition for offshell susy}
\end{align}
Here, $\epsilon$ is the one satisfying (\ref{five conditions for epsilon}).
The first two equations in (\ref{condition for offshell susy}) imply that 
$\nu^I (I=1,2, \cdots, 7)$ and $\Gamma^{0M} \epsilon (M=1,2, \cdots, 9)$ are 
orthonormal and the third equation is their completeness relation.

By using the bilinear forms found in the previous subsection, 
one can show that the equations (\ref{condition for offshell susy}) are solved by 
\begin{align}
\nu^I = \Gamma^{I9} \epsilon,
\label{def of nuI}
\end{align}
for $I= 1, 2, \cdots, 7$.


\section{Path integral of $U(1)$ modes}
\label{Path integral of U1 modes}

In this appendix, we perform the integration of overall $U(1)$ modes.
All fields are denoted by the same letters as the general $U(N)$ modes
but they shall indicate the $U(1)$ modes in this appendix.

The action (\ref{action of BFSS})
with the boundary term (\ref{def of boundary terms})
for the $U(1)$ modes are given by
\begin{align}
S^{U(1)}+S^{U(1)}_b = \frac{1}{g^2}
\int^{\tau_1}_{\tau_0} d\tau \; \left(
\frac{1}{2} (\partial X )(\partial \Phi) 
+\frac{1}{2} (\partial X^I)^2 -\frac{1}{2} K_I^2 -iv^0 \Psi_9 \partial \Psi_8
-iv^0 \Upsilon_I \partial \Psi_I
\right),
\label{U1 action with Sb}
\end{align}
where $X= X^8+X^0$ and $\Phi =X^8-X^0$.
Note that $K_I$ can be trivially integrated out and 
the $U(1)$ gauge field and ghosts are decoupled and only produce some 
numerical factor of the path integral. Thus, we ignore them below and only
consider the integration of the other fields.

For the path integral, we first specify the boundary conditions.
For simplicity, we first set the Dirichlet boundary conditions,
\begin{align}
&X^I (\tau_i) = x_i^I  \;\;\; (I=1, \cdots, 7),
\nonumber\\
&\Phi (\tau_i) = \phi_i,
\nonumber\\
&X (\tau_i)  = x_i,
\nonumber\\
&\Psi^M (\tau_i) = \eta_i^M  \;\;\; (M=1, 2, \cdots, 8),
\nonumber\\
&\partial \Psi^9 (\tau_i) = 0,
\nonumber\\
&\partial \Upsilon_I (\tau_i) =0,
\end{align}
for $i=0, 1$, where the objects on the right-hand sides are constants\footnote{
In the canonical formalism, 
the last two conditions can be understood as the equation of motion.}.

The transverse modes $X^I$ can be expanded as
\begin{align}
X^ I(\tau) = x_0^I + \frac{\tau -\tau_0}{\tau_1-\tau_0}(x_1^I -x_0^I) + 
\sum_{n=1}^\infty a_n^I \sin \left( 
\frac{\tau-\tau_0}{\tau_1-\tau_0} n\pi \right).
\label{expansion of XI}
\end{align}
By substituting this expansion, the $X^I$ part of the action is reduced to
\begin{align}
\int d\tau \frac{1}{2g^2} (\partial X^I)^2 
&=
\frac{(x_1^I -x_0^I)^2}{2g^2(\tau_1-\tau_0)} + \frac{1}{2g^2}
\sum_{n=1}^\infty a_n^I a_n^I \frac{n^2\pi^2}{2(\tau_1-\tau_0)}
\end{align}
The Gaussian integration of $a_n^I$ produces some numerical factors 
which do not depend on $x_i^I$. Thus, the integration of $X_I$ gives a 
factor $\exp\left(-\frac{(x_1^I -x_0^I)^2}{2g^2(\tau_1-\tau_0)} \right)$ to the partition function.

The fields $\Phi$ and $X$ can be integrated in a similar way.
The only difference is that after the similar expansion 
to (\ref{expansion of XI}), the action takes the 
following form, which is not Gaussian for each variables:
\begin{align}
\int d\tau \frac{1}{2g^2} (\partial X)(\partial \Phi) 
&=
\frac{(x_1 -x_0)(\phi_1-\phi_0)}{2g^2(\tau_1-\tau_0)} + \frac{1}{2g^2}
\sum_{n=1} a_n^X a_n^{\Phi} \frac{n^2\pi^2}{2(\tau_1-\tau_0)}
\end{align}
By making a Wick rotation for $a_n^\Phi$, the integration of 
$a_n^\Phi$ produces $\delta(a_n^{X})$. Then, $a_n^X$ can be 
trivially integrated out. Thus, these fields give a factor 
$\exp \left(-\frac{(x_1 -x_0)(\phi_1-\phi_0)}{2g^2(\tau_1-\tau_0)} \right)$
to the partition functions.

We next consider fermionic fields in (\ref{U1 action with Sb}).
Note that they all have the common form,
\begin{align}
\int D \Upsilon D\Psi \exp{ {\frac{iv^0}{g^2} \int_{\tau_0}^{\tau_1} d\tau \Upsilon \partial \Psi}},
\end{align}
with the boundary conditions $\Psi(\tau_i)=\eta_i$ and $\partial \Upsilon_I(\tau_i) =0$.
This type of integration turns out to produce the delta function, 
\begin{align}
\int D \Upsilon D\Psi \exp{ {\frac{iv^0}{g^2} \int_{\tau_0}^{\tau_1} d\tau \Upsilon \partial \Psi}} \propto \delta(\eta_1-\eta_0).
\end{align}
This is obtained as follows. We first expand $\Psi$ as
\begin{align}
\Psi(\tau)=\eta_0 + \frac{\tau-\tau_0}{\tau_1-\tau_0} (\eta_1-\eta_0) + \sum_{n=1}^\infty 
b_n \sin\left( 
\frac{\tau-\tau_0}{\tau_1-\tau_0} n\pi \right)
\end{align}
with the Grassmann variables $b_n$.
Note that, the constant mode of $\Upsilon$ couples to $\eta_1-\eta_0$.
By integrating this mode, we obtain the delta function 
$\delta(\eta_1-\eta_0)$. The other modes of $\Upsilon$ couple to $b_n$ and 
integration of these modes just produces a numerical factor.

Summarizing the above results, we find that for the 
Dirichlet boundary condition,  the partition function 
is given by 
\begin{align}
Z_{U(1)}^{\rm Dirichlet} =  \prod_{I=1}^8 \delta^8(\eta_1^I-\eta_0^I)
\exp \left\{  
-\frac{(x_1 -x_0)(\phi_1-\phi_0)}{2g^2(\tau_1-\tau_0)}
-\frac{(x_1^I -x_0^I)^2}{2g^2(\tau_1-\tau_0)}
\right\}
\label{ZU1 with Dirichlet}
\end{align}
up to an overall numerical factor.

The partition function with the Neumann boundary condition can be 
obtained from (\ref{ZU1 with Dirichlet}) by a Fourier transformation. 
For example, to change the boundary condition of $X^I$ (with fixed $I$), 
to the Neumann type, we perform the Fourier transformation,
\begin{align}
\int dx_0^I dx^I_1 e^{-\frac{(x^I_1-x^I_0)^2}{2g^2(\tau_1-\tau_0)}} e^{ix^I_1 p^I_1} e^{-ix^I_0 p^I_0}
& \propto 
\delta (p^I_1-p^I_0) e^{-\frac{g^2 (\tau_1-\tau_0)}{2} (p_1^I)^2}.
\end{align}
The variables $p^I_i$ correspond to the boundary values of
$\partial X^I$.
The transformation for $X$ and $\Phi$ is a little tricky. 
We first make a Wick rotation of $\phi_i$ and then make 
the Fourier transformation as
\begin{align}
\int dx_1 dx_0 d\phi_1 d\phi_0 
\exp \left( -i\frac{(x_1-x_0)(\phi_1-\phi_0)}{2(\tau_1-\tau_0)} 
\right) e^{ip^X_1 x_1 -ip^X_0x_0} e^{i p_1^\Phi \phi_1 -i p_0^\Phi \phi_0}.
\end{align}
By performing this integral, we obtain 
\begin{align}
 \delta(p_1^X-p^X_0) \delta(p_1^\Phi- p_0^\Phi) 
e^{-2i g^2(\tau_1-\tau_0) p_0^X p_0^\Phi}.
\end{align}
The partition function for other fields can also be transformed into the 
Neumann case in the same way\footnote{Note that the fermionic Fourier transformation of the fermionic delta function again gives a delta function.}.  
If we change the boundary condition of all the fields to the Neumann type,
we find the partition function,
\begin{align}
\tilde{Z}_{U(1)}^{\rm Neumann} 
= \delta^{(8)}(\tilde{\eta}_1^I -\tilde{\eta}_0^I)
\delta(p_1^I -p_0^I) \delta(p^X_1-p^X_0) \delta(p_1^\Phi-p_0^\Phi)
e^{-\frac{g^2(\tau_1-\tau_0)}{2}(p_1^I)^2  -2ig^2(\tau_1-\tau_0)p_1^X p_1^\Phi},
\end{align}
where $p^I_i, p_i^X, p_i^\Phi$ and $\tilde{\eta}_i^I$ are the boundary values of 
$\partial X^I, \partial X, \partial \Phi$ and $\partial \Psi^I$, respectively.

In order for the partition function to have 
non-trivial value in the limit 
$\tau_1 \rightarrow \infty$ and/or $\tau_0 \rightarrow -\infty$,  
$p^I_i$ must vanish and either $p^X_i$ or $p^\Phi_i$ also must vanish. 
However, physical variables are the rescaled quantities, $\tilde{p}_i^M=
 g^{4/3}p_i^M $, which have mass dimension 1.
Thus, we can take finite values of $\tilde{p}_i^M (\neq 0)$ by 
scaling the coupling constant as 
$g^{2/3} \gg (\tau_1 -\tau_0) (\tilde{p}_i^M)^2$.


\section{Some details of localization}

In this section, we show some details of the computation of localization.

\subsection{Solving saddle point equations}
\label{Solving saddle point equations}
Here, we solve the saddle point equations (\ref{saddle eq}) for $N=2$
with the boundary condition discussed in 
section \ref{Boundary condition}.

By using the values of bilinear forms shown in 
appendix \ref{Bilinear forms},
we can find that the second equation in (\ref{saddle eq}) is 
equivalent to the following 7 equations:
\begin{align}
&F_{91} -F_{23} -F_{47} +F_{56}=0,
\nonumber\\
&F_{92} -F_{31} -F_{46} -F_{57}=0,
\nonumber\\
&F_{93} -F_{12} +F_{45} -F_{67}=0,
\nonumber\\
&F_{94}+F_{17}+F_{26}-F_{35} =0,
\nonumber\\
&F_{95}-F_{16}+F_{27}+F_{34} =0,
\nonumber\\
&F_{96} +F_{15}-F_{24} +F_{37}=0,
\nonumber\\
&F_{97} -F_{14}-F_{25} -F_{36}=0.
\label{equation for X}
\end{align}
We take the $X_9=0$ gauge and put $D= \partial$ below.

We first solve (\ref{equation for X}) for $X_{4, 5, 6, 7}$. 
By differentiating the fourth equation of (\ref{equation for X}), we obtain
\begin{align}
\partial^2 X_4 -i \partial [X_1, X_7] -i \partial [X_2, X_6] +i \partial [X_3, X_5] =0.
\end{align}
Here, we can again use the other equations in (\ref{equation for X})
for the terms with a single derivative.
By using the Jacobi identity, 
the above equation is then reduced to
\begin{align}
(\partial^2 +  \sum_{I=4}^8 D_{I}^2)X_4 =0.
\end{align}
We then multiply $X_4$ from the left, take the trace and also integrate it over $\tau$.
Since $X_{4, 5, 6, 7}$ are assumed to be vanishing at the boundaries, 
we obtain
\begin{align}
 \int_{\tau_0}^{\tau_1} d\tau \; {\rm Tr}
\left( 
 (\partial X_4)^2 + \sum_{I=4}^7 (D_I X_4)^2
\right) =0.
\end{align}
Since the left-hand side is the sum of semi-positive terms,
each term should vanish. 
Then, we find that $\partial X_4 =0$ and $X_4$ is constant. 
Since the boundary condition is $X_4=0$, 
we find that $X_4= 0$ everywhere. 
We can apply the same argument to $X_{5, 6, 7}$ and obtain 
$X_{5, 6, 7} =0$ at the saddle point.

By putting $X_{4, 5, 6, 7}=0$ in (\ref{equation for X}), we obtain
\begin{align}
&\partial X_a + \frac{i}{2}\epsilon_{abc}[X_b, X_c] = 0,
\label{Nahm equation}
\end{align}
This is the so-called Nahm equation \cite{Nahm:1979yw}. 
The general solution to this equation for $\tau_1 = \infty$
with $N=2$ 
is constructed in \cite{Braden:2022ljy} and is given by 
$\hat{X}_a$ in (\ref{solution for N=2}).

We then solve the first equation of (\ref{saddle eq})
for $\tau_1 = \infty$ and $N=2$.
By using $X_{4, 5, 6, 7}=0$ and 
$\hat{X}_a$ in (\ref{solution for N=2}), the equation becomes
\begin{align}
\partial^2 G - \frac{c^2}{\sinh^2(c(\tau_0-\tau))}\sum_{i=1}^2[J_i^{[2]}, [J_i^{[2]}, G]]
-\frac{c^2}{\tanh^2(c(\tau_0-\tau))}[J_3^{[2]}, [J_3^{[2]}, G]] = 0,
\end{align}
where $G$ is either $\Phi'$ or $X$.
By expanding $G= G^a J_a^{[2]}$, it is easy to see that the above equation 
implies $G^1= G^2=0$.
Then, $g:=G^3$ satisfies
\begin{align}
\partial^2 g -\frac{2c^2}{\sinh^2(c(\tau_0-\tau))}  g=0.
\label{equation for g}
\end{align}
The general solution to this equation is given by
\begin{align}
g(\tau) =c_1 \frac{1}{\tanh(c(\tau_0-\tau))}
+ c_2 \left( 1- \frac{c(\tau_0-\tau)}{\tanh(c(\tau_0-\tau))} \right),
\label{divergent kernel}
\end{align}
where $c_1, c_2$ are the integration constants.
Since the term with $c_1$ is singular at $\tau = \tau_0$, we put $c_1 =0$.
For $G= \Phi'$ with the boundary condition $\partial \Phi'=0$, 
$c_2$ is also zero, while 
for $G= X$, $c_2$ is determined from (\ref{bc for X Phi})
as $c_2 =(p_3-p_2)/|c|$. Thus, we obtain the solution (\ref{solution for N=2}).


\subsection{Derivation of (\ref{1-loop det})}
\label{Derivation of 1-loop det}

In this appendix, we present a derivation of (\ref{1-loop det}).

We consider the integration 
\begin{align}
I(Z_0', Z_1') : = \int DZ_0 DZ_1  e^{-t Z_1' D_{10}Z_0 +t Z_1D_{10}Z_0'}
\label{def of I}
\end{align}
for $Z_0' \in ({\rm Ker}D_{10})^\perp$ and $Z_1' \in ({\rm Coker}D_{10})^\perp$,
where $V^\perp$ means the orthogonal complement of a vector space $V$ \footnote{
For a compact segment $[\tau_0, \tau_1]$, the operator $D_{10}$ is gapped and 
the separation between ${\rm Ker}D_{10}$ and $({\rm Ker}D_{10})^\perp$ is 
well-defined. We consider the case of half line $[\tau_0, \infty)$ as a limit of 
the segment as $\tau_1 \rightarrow \infty$. }.
We make Wick rotations for $C$ and $\Phi$ 
so that the exponent of (\ref{def of I}) becomes pure imaginary.
The expression (\ref{1-loop det}) is derived if we have
\begin{align}
I(Z_0', Z_1') = \delta(Z_0')\delta(Z_1'),
\label{result for I}
\end{align}
for $Z_0' \in ({\rm Ker}D_{10})^\perp$ and $Z_1' \in ({\rm Coker}D_{10})^\perp$.
We will show (\ref{result for I}) in the following.

Note first that if $D_{10}$ has nontrivial kernel or cokernel, 
the integral (\ref{def of I}) may not be well-defined.
However, as we show  
in appendices \ref{computation of ker} and \ref{computation of coker D10},
this is not the case.
The configuration spaces of $Z_0$ and $Z_1$
do not have nontrivial kernel and cokernel of $D_{10}$, respectively, 
and the integral (\ref{def of I}) is well-defined.

Let $H_0$ and $H_1'$ be the Hilbert spaces of the bosonic fields 
$Z_0$ and $Z_1'$, respectively.
The operator $D_{10}: H_0 \rightarrow H_1'$ is closed\footnote{
Though $D_{10}$ contains some Nahm poles, by introducing a cutoff near the 
boundary, it becomes a closed operator.} and thus gives an isomorphism 
$H_0/{\rm Ker}D_{10} \cong H_1'/{\rm Coker}D_{10}$. 
As shown in appendices \ref{computation of ker} and \ref{computation of coker D10}, the spaces ${\rm Ker} D_{10} \subset H_0$
and ${\rm Coker} D_{10} \subset H_1'$ are both trivial. Then, we have 
the isomorphism,
\begin{align}
H_0 \cong H_1'.
\label{iso of H0 and cokerperp}
\end{align}
Thus, in these spaces, it makes sense to consider the eigenvalues
or the determinant of $D_{10}$. 
Since the operator $D_{10}$ 
(after the Wick rotation of $C$ and $\Phi$) preserves the 
Hermiticity of the matrices, its eigenvalues and determinant should be real.
Then, the integration over $Z_0$ in (\ref{def of I}) is computed as
\begin{align}
\int DZ_0 e^{-t Z_1' D_{10} Z_0} 
\propto \int DW_1  \frac{1}{{\rm det}(tD_{10})} e^{iZ_1' W_1} 
= \frac{1}{{\rm det}(tD_{10})} \delta(Z_1')
\label{det D in bosons}
\end{align}
for $Z_1' \in ({\rm Coker}D_{10})^\perp$, where we changed the variable
from $Z_0$ to $W_1 = i t D_{10}Z_0 \in  ({\rm Coker}D_{10})^\perp$ and dropped an 
overall constant phase.

Similarly, let $H_0'$ and $H_1$ be the Hilbert spaces of the fermionic fields
$Z_0'$ and $Z_1$, respectively.
The operator $D_{10}: H_0' \rightarrow H_1$ gives
the isomorphism $H_0'/{\rm Ker}D_{10} \cong H_1/{\rm Coker}D_{10}$.
It is shown in appendices \ref{computation of ker} and \ref{computation of coker D10} that 
while ${\rm Coker}D_{10} \subset H_1$ is trivial, 
${\rm Ker}D_{10} \subset H_0'$ is not. Then, we have the 
isomorphism
\begin{align}
H_0' \supset ({\rm Ker}D_{10})^{\perp} \cong H_1.
\end{align}
The integration over $Z_1$ in (\ref{def of I}) is then computed as
\begin{align}
\int DZ_1 e^{tZ_1 D_{10} Z_0'}
=\delta(tD_{10} Z_0') = {\rm det}(tD_{10}) \delta(Z_0')
\label{det D in fermions}
\end{align}
for $Z_0' \in ({\rm Ker}D_{10})^\perp$.

Note that the two determinants in (\ref{det D in bosons}) and (\ref{det D in fermions})
are taken in different Hilbert spaces. One is taken in $H_1'$ and the other is in $H_1$.
However, there is another isomorphism given by the 
supersymmetry $Q$.
Since $Z_1$ and $Z_1'$ have exactly the same boundary conditions, we have the obvious isomorphism $H_1 \cong H_1'$
given by (\ref{eq for Q-complex}).
Hence, we can identify the two determinants. 
Therefore, from (\ref{det D in bosons}) 
and (\ref{det D in fermions}), we obtain (\ref{result for I}).

\subsection{Computation of ${\rm Ker}D_{10}$}
\label{computation of ker}

In this subsection, we compute ${\rm Ker}D_{10}$.

Let $H_i$ and $H_i'$ be the Hilbert space of $Z_i$ and $Z_i'$,
respectively, for $i=0,1$.
The operator $D_{10}$ acts on the bosonic Hilbert space as 
$H_0 \rightarrow H_1'$ 
and the fermionic Hilbert space as $H_0' \rightarrow H_1$.
We first compute kernel in the fermionic space and then 
consider the bosonic case later.

\subsubsection{${\rm Ker}D_{10} \subset H_0'$}

For $\tilde{\Psi}_{M'} \in {\rm Ker}D_{10}$, each coefficient of $C$, $\tilde{C}$ and
$\Upsilon_I$ is vanishing in (\ref{Z1D10Z0'}).
We first solve the equations coming from the coefficients of $\Upsilon_{4, 5, 6, 7}$,
\begin{align}
\partial \tilde{\Psi}_I + i \frac{w^{09IaJ}}{v^0} [\hat{X}_a, \tilde{\Psi}_J] =0
\;\;\;\; (I=4, 5, 6, 7).
\label{equation for upsilon 4567}
\end{align}
Let us define the first order differential operator
\begin{align}
\Delta = 
\left(
\begin{array}{cccc}
\partial & -\hat{D}_3 & \hat{D}_2 & \hat{D}_1 \\
\hat{D}_3 & \partial & -\hat{D}_1 & \hat{D}_2 \\
-\hat{D}_2 & \hat{D}_1 & \partial & \hat{D}_3 \\
-\hat{D}_1 & -\hat{D}_2 & -\hat{D}_3 & \partial \\
\end{array}
\right)
\label{def of Delta}
\end{align}
acting on $(\tilde{\Psi}_4, \tilde{\Psi}_5, \tilde{\Psi}_6, \tilde{\Psi}_7)^T$,
where $\hat{D}_a = -i[\hat{X}_a, \;\;]$. Then, 
by using (\ref{values of w}), we can write 
the equations 
(\ref{equation for upsilon 4567}) as
\begin{align}
(\Delta \tilde{\Psi})_I =0   \;\;\;\; (I=4,5,6,7).
\label{ equation for tilde Psi 4567}
\end{align}
We also define 
\begin{align}
\tilde{\Delta} = 
\left(
\begin{array}{cccc}
-\partial & -\hat{D}_3 & \hat{D}_2 & \hat{D}_1 \\
\hat{D}_3 & -\partial & -\hat{D}_1 & \hat{D}_2 \\
-\hat{D}_2 & \hat{D}_1 & -\partial & \hat{D}_3 \\
-\hat{D}_1 & -\hat{D}_2 & -\hat{D}_3 & -\partial \\
\end{array}
\right).
\label{def of Delta tilde}
\end{align}
By using the fact that 
$\hat{X}_a$ satisfy the Nahm equation (\ref{Nahm equation}), 
we find that $\Delta$ and $\tilde{\Delta}$ satisfy
\begin{align}
\tilde{\Delta} \Delta = -(\partial^2 +\hat{D}_a^2) {\bf 1}_{4}.
\label{tilde Delta Delta is positive}
\end{align}
Note that the right-hand side is positive semi definite. 
This implies that the equation
(\ref{ equation for tilde Psi 4567}) only has the trivial solution
$\tilde{\Psi}_I(\tau) =0$.

We next consider the coefficients of $\Upsilon_{a}$ and $\tilde{C}$
in (\ref{Z1D10Z0'}). 
By putting $\tilde{\Psi}_I =0 (I=4,5,6,7)$ in these coefficients, 
and equating each of the coefficients to zero, we find the following equations:
\begin{align}
&\partial \tilde{\Psi}_a + i[\hat{X}_a, \tilde{\Psi}_9] + i\epsilon_{abc}
[\hat{X}_b, \tilde{\Psi}_c] = 0,
\nonumber\\
&\partial \tilde{\Psi}_9 -i[\hat{X}_a, \tilde{\Psi}_a] =0.
\label{Ker equations for psia and psi9}
\end{align}
Note that at the boundary, the first equation is the same 
as (\ref{bc for fermions}), so that any finite solution of 
the above equations satisfies the boundary condition.
The equations (\ref{Ker equations for psia and psi9})
can also be written as
\begin{align}
(\tilde{\Delta} \tilde{\Psi})_M =0  \;\;\;\; (M=1,2,3,9),
\label{zero mode equation of Delta tilde}
\end{align}
where we regard $\tilde{\Delta}$ as an operator acting 
on $(\tilde{\Psi}_1, \tilde{\Psi}_2, \tilde{\Psi}_3, \tilde{\Psi}_9)^T$.
Below, we will show that $\tilde{\Delta}$ has nontrivial kernel.

\subsubsection*{Angular momentum decomposition}

We first note that $\tilde{\Delta}$ can be written as 
\begin{align}
\tilde{\Delta} = -\partial -2is_a \hat{D}_a.
\end{align}
Here, $s_a$ are given by
\begin{align}
&s_1=-\frac{1}{2} \sigma^1 \otimes \sigma^2,
\nonumber\\
&s_2=-\frac{1}{2} \sigma^2 \otimes {\bf 1}_2,
\nonumber\\
&s_3=\frac{1}{2} \sigma^3 \otimes \sigma^2,
\end{align}
where $\sigma^a$ are the Pauli matrices.
$s_a$ satisfy the standard $SU(2)$ Lie algebra, 
$[s_a, s_b]= i\epsilon_{abc} s_c$. 
The representation of $s_a$ is four dimensional and its second Casimir is 
given by $s_a s_a = \frac{3}{4}{\bf 1}_4$. Thus, this representation consists of 
two copies of the spin $1/2$ representation of $SU(2)$ Lie algebra, namely, 
$s_a \cong J^{[2]}_a \oplus J^{[2]}_a$ up to a unitary similarity transformation.
An orthonormal basis of this representation space is explicitly  given by
\begin{align}
&| -1/2, 1 \rangle = 
\frac{1}{\sqrt{2}}
\left(
\begin{array}{c}
1 \\
-i \\
0 \\
0 \\
\end{array}
\right),\;\;\;\;\;
| 1/2, 1 \rangle = 
\frac{1}{\sqrt{2}}
\left(
\begin{array}{c}
0 \\
0 \\
1 \\
-i \\
\end{array}
\right),
\nonumber\\
&| -1/2, 2 \rangle = 
\frac{1}{\sqrt{2}}
\left(
\begin{array}{c}
0 \\
0 \\
1 \\
i \\
\end{array}
\right),\;\;\;\;\;
| 1/2, 2 \rangle = 
\frac{1}{\sqrt{2}}
\left(
\begin{array}{c}
-1 \\
-i \\
0 \\
0 \\
\end{array}
\right),
\end{align}
where we denote by $|s, \alpha \rangle $ the eigenstate of $s_3$ with eigenvalue $s$ 
belonging to the $\alpha$th copy of $J_a^{[2]}$.
The relative phase for $s=\pm 1/2$ is fixed in a way that the standard relation
$s_+ | -1/2, \alpha \rangle = | 1/2, \alpha \rangle$ is satisfied, where 
$s_+ = s_1 +i s_2$.

We also note that since $\hat{X}_a$ in
(\ref{solution for N=2}) are proportional to $J_a^{[2]} = \sigma_a/2$, 
the operators $\hat{D}_a = -i[\hat{X}_a, \;\; ]$ include the adjoint action of $J_a^{[2]}$, 
which we denote by ${\rm Ad}(J_a^{[2]}) = [J_a^{[2]},  \;\; ]$.
This acts on the $2\times 2$ matrices, which can be decomposed to 
a triplet and a singlet under the adjoint action. The singlet is the overall 
$U(1)$ modes proportional to the identity matrix, which we do not consider 
here. The triplet is the $SU(2)$ modes. We regard these modes as
the spin $1$ representation. 
An orthonormal basis of this representation is given by
\begin{align}
| 1 \rangle = J_+^{[2]},  \;\;\;  | 0 \rangle = -\sqrt{2} J_3^{[2]},  
\;\;\;  |-1 \rangle = -J_-^{[2]},
\end{align}
where we denote by $|n \rangle $ the eigenstate of ${\rm Ad}(J_a^{[2]})$.
We again fixed the relative phase by demanding 
${\rm Ad}(J_-^{[2]}) |n \rangle = \sqrt{(1+n)(2-n)} | n-1 \rangle$.
The overall normalization is fixed by $1=\langle 1 | 1 \rangle = {\rm Tr}(J_-^{[2]}J_+^{[2]})$ and so on.

If we write ${\rm Ad}(J_a^{[2]}) = J_a^{[3]}$, we can write 
\begin{align}
&\hat{D}_a  =-\frac{ic}{\sinh(c(\tau_0-\tau))} J_a^{[3]} \;\;\; (a=1,2),
\nonumber\\
&\hat{D}_3 =  -\frac{ic}{\tanh(c(\tau_0-\tau))} J_3^{[3]}.
\end{align}
In this notation, we have
\begin{align}
\tilde{\Delta} = -\partial -\frac{2c}{\sinh(c(\tau_0-\tau))} s_a J_a^{[3]}
-\frac{2c(\cosh(c(\tau_0-\tau))-1)}{\sinh(c(\tau_0-\tau))}s_3 J_3^{[3]}.
\end{align}
The field $\tilde{\Psi}$ in the eigenvalue problem 
(\ref{zero mode equation of Delta tilde}) can be expanded as
\begin{align}
\tilde{\Psi}(\tau) = \sum_{J=1/2}^{3/2} \sum_{\alpha=1}^2 \sum_{s=-1/2}^{1/2} \sum_{n=-1}^1 \Psi_{J, s+n}^\alpha(\tau) C^{J s+n}_{\frac{1}{2}s, 1 n}      |s, \alpha \rangle  | n \rangle,
\label{expantion of psi}
\end{align}
where $C^{J s+n}_{\frac{1}{2}s, 1 n}$ is the Clebsch-Gordan coefficient.
The equation (\ref{zero mode equation of Delta tilde}) is then equivalent to 
\begin{align}
\sum_{J=1/2}^{3/2} 
C^{J s+n}_{\frac{1}{2}s, 1 n}
\left[ 
\partial 
+ \frac{2c \left(\frac{J(J+1)}{2}-\frac{11}{8} \right)}{\sinh T}
+ \frac{2c (\cosh T-1)}{\sinh T }sn
\right]\Psi_{J, s+n}^\alpha =0
\label{equation decomposed with angular momenta}
\end{align}
for $\alpha =1,2$, $s=\pm1/2$ and $n=-1,0,1$. Here, we 
defined $T=c(\tau_0-\tau)$ and used
the elementary identity, 
$s_a J_a^{[3]} = \frac{J(J+1)}{2} -\frac{1}{2} s_a^2 -\frac{1}{2} (J_a^{[3]})^2$
which holds for states with total angular momentum $J$.

We list the values of the Clebsch-Gordan coefficients that  we will use
for solving (\ref{equation decomposed with angular momenta}).
\begin{align}
&C^{\frac{3}{2}\frac{1}{2}}_{\frac{1}{2}\frac{1}{2}, 10} = \sqrt{\frac{2}{3}},  \;\;\;\;\;\;
C^{\frac{3}{2}\frac{1}{2}}_{\frac{1}{2}-\frac{1}{2}, 11} = \frac{1}{\sqrt{3}},  \;\;\;\;\;\;
C^{\frac{3}{2}-\frac{1}{2}}_{\frac{1}{2}-\frac{1}{2}, 10} =\sqrt{\frac{2}{3}},  \;\;\;\;\;\;
C^{\frac{3}{2}-\frac{1}{2}}_{\frac{1}{2}\frac{1}{2}, 1-1} =\frac{1}{\sqrt{3}}, 
\nonumber\\
&C^{\frac{1}{2}\frac{1}{2}}_{\frac{1}{2}\frac{1}{2}, 10} = \frac{1}{\sqrt{3}}, \;\;\;\;\;\;
C^{\frac{1}{2}\frac{1}{2}}_{\frac{1}{2}-\frac{1}{2}, 11} = -\sqrt{\frac{2}{3}},  \;\;\;\;\;\;
C^{\frac{1}{2}-\frac{1}{2}}_{\frac{1}{2}-\frac{1}{2}, 10} = -\frac{1}{\sqrt{3}},  \;\;\;\;\;\;
C^{\frac{1}{2}-\frac{1}{2}}_{\frac{1}{2}\frac{1}{2}, 1-1} = \sqrt{\frac{2}{3}}.
\end{align}

\subsubsection*{$ \bullet \; (s, n)=(1/2, 1), (-1/2, -1)$}
For $(s, n)=(1/2, 1)$, the Clebsch-Gordan coefficient is vanishing 
unless $J=3/2$. Thus, 
(\ref{equation decomposed with angular momenta}) is reduced to
\begin{align}
\left[ 
\partial 
+ \frac{c \cosh T}{\sinh T}
\right] \Psi^\alpha_{\frac{3}{2}, \frac{3}{2}}=0.
\end{align}
The solution to this equation is given by 
\begin{align}
\Psi^\alpha_{\frac{3}{2}, \frac{3}{2}}
=A^\alpha \sinh T
\end{align}
with integration constant $A^\alpha$.
Since this function is divergent as $\tau \rightarrow \infty$ and does not 
satisfies the boundary condition, 
we set $A^\alpha =0$.
Then we only have the trivial solution 
$\Psi^\alpha_{\frac{3}{2}, \frac{3}{2}}=0$ in this case.

The case with $(s, n) =(-1/2, -1)$ can be treated in the same way 
and we obtain $\Psi^\alpha_{\frac{3}{2}, -\frac{3}{2}}=0$.
  
\subsubsection*{$\bullet \; (s, n)=(1/2, 0), (-1/2, 1)$}
For $(s, n) =(1/2, 0)$ and $(s, n) =(-1/2, 1)$, the equation 
(\ref{equation decomposed with angular momenta})
reduces to
\begin{align}
&\sqrt{2}
\left(
\partial +\frac{c}{\sinh T}
\right) \Psi_{\frac{3}{2}, \frac{1}{2}}^\alpha
+ 
\left(
\partial -\frac{2c}{\sinh T}
\right) \Psi_{\frac{1}{2}, \frac{1}{2}}^\alpha =0, 
\label{eq1}
\end{align}
and 
\begin{align}
\left(
\partial -\frac{c(\cosh T -2 )}{\sinh T}
\right) \Psi_{\frac{3}{2}, \frac{1}{2}}^\alpha
-\sqrt{2}
\left(
\partial -\frac{c(\cosh T +1 )}{\sinh T}
\right) \Psi_{\frac{1}{2}, \frac{1}{2}}^\alpha =0,
\label{eq2}
\end{align}
respectively. 
The general solution to these equations is 
\begin{align}
&\Psi^\alpha_{\frac{3}{2}, \frac{1}{2}}
= A^\alpha \frac{\sqrt{2}(1-\cosh T)}{\sinh^2 T}
+ B^\alpha \frac{2\sqrt{2}(T+\sinh T)
(1-\cosh T)}{\sinh^2 T},
\nonumber\\
&\Psi^\alpha_{\frac{1}{2}, \frac{1}{2}}
= A^\alpha \frac{1+2\cosh T}{\sinh^2 T}
+ B^\alpha \frac{2T-\sinh 2T +4T \cosh T -4 \sinh T}{\sinh^2 T},
\label{solution for m=1/2}
\end{align}
where $A^\alpha$ and $B^\alpha$ are integration constants.
Since the first term of the second equation diverges as $\tau \rightarrow \tau_0$, we set $A^\alpha =0$. 
Note that the terms with $B^\alpha$ 
are convergent both in $\tau \rightarrow \tau_0$ and $\tau \rightarrow \infty$.

\subsubsection*{$\bullet \; (s, n)=(-1/2, 0), (1/2, -1)$}
For $(s, n) =(-1/2, 0)$ and $(s, n) =(1/2, -1)$, the equation 
(\ref{equation decomposed with angular momenta})
reduces to
\begin{align}
&\sqrt{2}
\left(
\partial +\frac{c}{\sinh T}
\right) \Psi_{\frac{3}{2}, -\frac{1}{2}}^\alpha
- 
\left(
\partial -\frac{2c}{\sinh T}
\right) \Psi_{\frac{1}{2}, -\frac{1}{2}}^\alpha =0, 
\end{align}
and 
\begin{align}
\left(
\partial -\frac{c(\cosh T -2 )}{\sinh T}
\right) \Psi_{\frac{3}{2}, -\frac{1}{2}}^\alpha
+\sqrt{2}
\left(
\partial -\frac{c(\cosh T +1 )}{\sinh T}
\right) \Psi_{\frac{1}{2}, -\frac{1}{2}}^\alpha =0,
\end{align}
respectively. These are the same form as 
(\ref{eq1}) and (\ref{eq2}) under the replacements 
$(\Psi_{\frac{3}{2}, -\frac{1}{2}}^\alpha, \Psi_{\frac{1}{2}, -\frac{1}{2}}^\alpha)
\rightarrow 
(\Psi_{\frac{3}{2}, \frac{1}{2}}^\alpha, -\Psi_{\frac{1}{2}, \frac{1}{2}}^\alpha)$.
Thus, the solution is given by
\begin{align}
&\Psi^\alpha_{\frac{3}{2}, -\frac{1}{2}}
= C^\alpha \frac{2\sqrt{2}(T+\sinh T)
(1-\cosh T)}{\sinh^2 T},
\nonumber\\
&\Psi^\alpha_{\frac{1}{2}, -\frac{1}{2}}
= -C^\alpha \frac{2T-\sinh 2T +4T \cosh T -4 \sinh T}{\sinh^2 T}.
\label{solution for m=-1/2}
\end{align}

\subsubsection*{Total solution}

By substituting the solutions 
(\ref{solution for m=1/2}) with $A^\alpha=0$ and (\ref{solution for m=-1/2}) into 
(\ref{expantion of psi}), we obtain
\begin{align}
\tilde{\Psi} = \sqrt{3}
\left(
\begin{array}{c}
(B^2-C^1) h_1 J_3^{[2]} + h_2(B^1 J_+^{[2]} +C^2 J_-^{[2]}) \\
i(B^2+C^1) h_1J_3^{[2]} -i h_2(B^1 J_+^{[2]} -C^2J_-^{[2]}) \\
-(B^1+C^2) h_1J_3^{[2]} +h_2(B^2 J_+^{[2]} -C^1 J_-^{[2]}) \\
i(B^1 -C^2) h_1J_3^{[2]} +ih_2(B^2J_+^{[2]} +C^1 J_-^{[2]}) \\
\end{array}
\right),
\end{align}
where we defined the functions $h_1(\tau)$ and $h_2(\tau)$ by
\begin{align}
h_1(\tau ) = \frac{2T-\sinh 2T}{ \sinh^2 T}, \;\;\;\;
h_2(\tau) = \frac{2(\sinh T -T \cosh T)}{\sinh^2 T}.
\end{align}

Now we impose the reality condition for $\Psi$, which implies that
$B^2= -(C^1)^*$ and $B^1 =(C^2)^*$. 
Then, by decomposing $B^1 = a-ib, B^2=c-id, C^1=-c-id, C^2 = a+ib$, 
and absorbing some constants by properly rescaling $a,b,c,d$, we obtain 
\begin{align}
\left(
\begin{array}{c}
\tilde{\Psi}_1 \\
\tilde{\Psi}_2 \\
\tilde{\Psi}_3 \\
\tilde{\Psi}_9 \\
\end{array}
\right)
&=
a \left(
\begin{array}{c}
 h_2 J_1^{[2]} \\
 h_2 J_2^{[2]} \\
-h_1 J_3^{[2]} \\
0   \\
\end{array}
\right)
+
b \left(
\begin{array}{c}
 h_2 J_2^{[2]} \\
- h_2 J_1^{[2]} \\
0 \\
h_1 J_3^{[2]}  \\
\end{array}
\right)
+
c \left(
\begin{array}{c}
h_1J_3^{[2]} \\
0 \\
h_2 J_1^{[2]} \\
-h_2 J_2^{[2]} \\
\end{array}
\right)
+
d \left(
\begin{array}{c}
0 \\
h_1J_3^{[2]} \\
h_2 J_2^{[2]} \\
h_2 J_1^{[2]}  \\
\end{array}
\right).
\label{zero mode fermions}
\end{align}

We finally consider the coefficient of $C$ in (\ref{Z1D10Z0'}). 
By substituting the solution (\ref{zero mode fermions}) and 
equating the coefficient to zero, we obtain 
\begin{align}
(\partial^2  +\hat{D}_a^2 )\tilde{\Psi}_8 =0.
\label{Ker equation for psi8}
\end{align}
This has only the trivial solution for the boundary condition
(\ref{bc for fermions}).

Thus, we finally find that
${\rm Ker}D_{10}$ consists of the 
four real zero modes corresponding to $a, b, c$ and $d$ 
given in (\ref{zero mode fermions}).

\subsubsection{${\rm Ker}D_{10} \subset H_0$}

We next consider kernel of $D_{10}$ in the bosonic Hilbert space.
The structure of $D_{10}$ is common both for bosons and fermions 
and the only difference comes from the boundary conditions.

For $X_{4,5,6,7}$, the boundary condition is 
the same as $\tilde{\Psi}_{4,5,6,7}$, so that we have the same conclusion 
as in the previous section that the kernel of $D_{10}$ is trivial for 
the space of $X_{4,5,6,7}$.

For $X_{1,2,3}, X_8$ and $X_9$, we also have the same conditions as 
(\ref{Ker equations for psia and psi9}) or (\ref{Ker equation for psi8}).
The second equation of (\ref{Ker equations for psia and psi9})
for $X_{1,2,3}$ and $X_9$ corresponds to the gauge fixing condition and 
is already taken into account in the 
BRST formalism. The first equation 
of (\ref{Ker equations for psia and psi9})
corresponds to the linear approximation of the Nahm equation around
the saddle-point configuration.
Since our saddle-point configuration is unique and does not allow any 
continuous moduli, we conclude that the space of $X_{1, 2, 3}$ and $X_9$ 
has only the trivial kernel. Similarly, the counterpart of 
(\ref{Ker equation for psi8}) for $X_8$ also does not allow any nontrivial 
solution.

Therefore, we conclude that ${\rm Ker}D_{10} \subset H_0$ is trivial.

\subsection{Computation of ${\rm Coker}D_{10}$}
\label{computation of coker D10}

In this subsection, we compute 
${\rm Coker}D_{10} \subset H_1'$ in the bosonic space and
show that ${\rm Coker}D_{10}=\{0\}$.
The cokernel in the fermionic space 
${\rm Coker}D_{10} \subset H_1$ should have exactly the same 
structure as the bosonic case, since the boundary conditions 
for $Z_1$ and $Z_1'$ are the same. 

The statement ${\rm Coker}D_{10}=\{0\}$ is equivalent to 
${\rm Ker}D_{10}D_{10}^\dagger =\{0\}$, where 
$D_{10}^{\dagger}$ is the adjoint of $D_{10}$.
We will show the latter statement in the following, namely, 
we will show that if we assume
\begin{align}
D_{10} D_{10}^\dagger  Z_1' =0,
\label{assumptions}
\end{align}
then, we have $Z_1'=0$. 

Note that if we arrange the fields as
\begin{align}
Z_1' = 
\left(
\begin{array}{c}
\tilde{H}_a \\
B \\
\tilde{\Phi} \\
\tilde{H}_I \\
\end{array}
\right), 
\;\;\;\; 
Z_0 =
\left(
\begin{array}{c}
X_a \\
X_9 \\
X_8 \\
X_I \\
\end{array}
\right), 
\end{align}
with $a =1, 2, 3$ and $I=4,5,6,7$, then $D_{10}: H_0 \rightarrow H_1'$ 
and $D_{10}^\dagger: H_1' \rightarrow H_0$ have 
the following structure
\begin{align}
D_{10} = 
\left(
\begin{array}{cc}
\tilde{\Delta}  &  0  \\
 F & G \\  
\end{array}
\right),
\;\;\;\;\;
D_{10}^\dagger = 
\left(
\begin{array}{cc}
\Delta  &  F^\dagger  \\
 0 & G^\dagger \\  
\end{array}
\right),
\end{align}
where $\Delta$ and $\tilde{\Delta}$ are defined in 
(\ref{def of Delta}) and (\ref{def of Delta tilde}), respectively, 
and, $F$ and $G$ are some operators. 
Here, we ignore the boundary terms in $D_{10}^\dagger$;
we first solve (\ref{assumptions}) in the bulk and 
discuss the effect of the boundary terms later.

From the assumption (\ref{assumptions}), we have
$D_{10}^\dagger  Z_1' \in {\rm Ker}D_{10}$. As shown in 
appendix \ref{computation of ker}, this implies that 
\begin{align}
D_{10}^\dagger  Z_1'
= \left(
\begin{array}{c}
0 \\
0 \\
0 \\
0 \\
\end{array}
\right).
\label{D dagger is in kernel D}
\end{align}
Then, this implies that 
\begin{align}
G^\dagger 
\left(
\begin{array}{c}
\tilde{\Phi} \\
\tilde{H_I} \\
\end{array}
\right) =0.
\label{ker G dagger is trivial}
\end{align}
Now, suppose that $G^\dagger$ only has the trivial kernel and the above
equation implies $\tilde{\Phi} =\tilde{H}_I = 0\; (I=4,5,6,7)$. 
Then, (\ref{assumptions}) reduces to 
\begin{align}
\tilde{\Delta} \Delta  
\left( 
\begin{array}{c}
\tilde{H}_a \\
 B  \\
\end{array}
\right) = 0.
\label{C48}
\end{align}
As shown in (\ref{tilde Delta Delta is positive}), this operator does not have 
nontrivial kernel\footnote{
The solution to (\ref{C48}) is given by (\ref{divergent kernel}), which
is divergent at $\tau = \tau_0$ or $\tau=\infty$. 
We do not include such divergent configuration in the path integral over
$B$, since $QV_{gh}$ diverges for such configuration because of 
the term $\frac{\xi}{2}B^2$.
}, so that $B=\tilde{H}_a =0$ and we obtain $Z_1'=0$.
Thus, it suffices to show that ${\rm Ker}G^\dagger = \{0\}$.

The equation (\ref{ker G dagger is trivial}) is explicitly written as
\begin{align}
&\partial^2 \tilde{\Phi} - [\hat{X}_a, [\hat{X}_a, \tilde{\Phi}]] = 0,
\nonumber\\
&\partial \tilde{H}_I + i \frac{w^{09a IJ}}{v^0} [\hat{X}_a, \tilde{H}_J] =0.
\end{align}
The latter equation can be rearranged into the form $(\Delta \tilde{H})_I =0$ and 
since the kernel of $\Delta$ is trivial, we obtain $\tilde{H}_I =0\; (I=4,5,6,7)$.
The former equation is identical to the saddle point equation of $\Phi$ and 
we already know that the only solution is given by $\tilde{\Phi}=0$. 
Thus, we find that ${\rm Ker}G^\dagger = \{0\}$
and the equation (\ref{assumptions}) only has the trivial solution 
$Z_1'=0$ in the bulk.

Because of the continuity of the fields, the conclusion is not altered
even after taking the boundary terms in $D_{10}^\dagger$ into account.
Therefore, ${\rm Coker}D_{10}=\{0\}$.

\end{appendix}


\begin{thebibliography}{99}

\bibitem{Banks:1996vh}
T.~Banks, W.~Fischler, S.~H.~Shenker and L.~Susskind,
Phys. Rev. D \textbf{55}, 5112-5128 (1997).


\bibitem{Polchinski:1997pz}
J.~Polchinski and P.~Pouliot,
Phys. Rev. D \textbf{56}, 6601-6606 (1997).


\bibitem{Herderschee:2023pza}
A.~Herderschee and J.~Maldacena,
J. Phys. A \textbf{57}, no.16, 165401 (2024).

\bibitem{Herderschee:2023bnc}
A.~Herderschee and J.~Maldacena,
JHEP \textbf{11}, 052 (2024).

\bibitem{Laurenzano:2025ywy}
D.~Laurenzano and J.~F.~Wheater,
[arXiv:2510.15488 [hep-th]].

\bibitem{Dijkgraaf:1997vv}
R.~Dijkgraaf, E.~P.~Verlinde and H.~L.~Verlinde,
Nucl. Phys. B \textbf{500}, 43-61 (1997).

\bibitem{Taylor:1996ik}
W.~Taylor,
Phys. Lett. B \textbf{394}, 283-287 (1997).

\bibitem{Pestun:2007rz}
V.~Pestun,
Commun. Math. Phys. \textbf{313}, 71-129 (2012).


\bibitem{Nahm:1979yw}
W.~Nahm,
Phys. Lett. B \textbf{90}, 413-414 (1980).


\bibitem{Braden:2022ljy}
H.~W.~Braden, S.~A.~Cherkis and J.~M.~Quinones,
J. Math. Phys. \textbf{64}, no.1, 011701 (2023).


\bibitem{Susskind:1997cw}
L.~Susskind,
[arXiv:hep-th/9704080 [hep-th]].

\bibitem{Sen:1997we}
A.~Sen,
Adv. Theor. Math. Phys. \textbf{2}, 51-59 (1998).

\bibitem{Seiberg:1997ad}
N.~Seiberg,
Phys. Rev. Lett. \textbf{79}, 3577-3580 (1997).


\bibitem{Diaconescu:1996rk}
D.~E.~Diaconescu,
Nucl. Phys. B \textbf{503}, 220-238 (1997).

\bibitem{Myers:1999ps}
R.~C.~Myers,
JHEP \textbf{12} (1999), 022.





\bibitem{Bachas:2000dx}
C.~Bachas, J.~Hoppe and B.~Pioline,
JHEP \textbf{07}, 041 (2001).

\bibitem{Yee:2003ge}
J.~T.~Yee and P.~Yi,
JHEP \textbf{02}, 040 (2003).

\bibitem{Kovacs:2015xha}
S.~Kovacs, Y.~Sato and H.~Shimada,
JHEP \textbf{02}, 050 (2016).



\bibitem{Bergshoeff:2018yvt}
E.~Bergshoeff, J.~Gomis and Z.~Yan,
JHEP \textbf{11}, 133 (2018).

\bibitem{Maskalaniec:2026vlk}
D.~Maskalaniec, Z.~Yan and U.~Zorba,
[arXiv:2603.10278 [hep-th]].


\bibitem{Berenstein:2002jq}
D.~E.~Berenstein, J.~M.~Maldacena and H.~S.~Nastase,
JHEP \textbf{04}, 013 (2002)
doi:10.1088/1126-6708/2002/04/013
[arXiv:hep-th/0202021 [hep-th]].


\bibitem{Asano:2012zt}
Y.~Asano, G.~Ishiki, T.~Okada and S.~Shimasaki,
JHEP \textbf{02}, 148 (2013).

\bibitem{Asano:2017nxw}
Y.~Asano, G.~Ishiki, S.~Shimasaki and S.~Terashima,
JHEP \textbf{02}, 076 (2018).


\bibitem{Asano:2014vba}
Y.~Asano, G.~Ishiki, T.~Okada and S.~Shimasaki,
JHEP \textbf{05}, 075 (2014)
doi:10.1007/JHEP05(2014)075
[arXiv:1401.5079 [hep-th]].

\bibitem{Asano:2014eca}
Y.~Asano, G.~Ishiki and S.~Shimasaki,
JHEP \textbf{09}, 137 (2014)
doi:10.1007/JHEP09(2014)137
[arXiv:1406.1337 [hep-th]].

\bibitem{Asano:2017xiy}
Y.~Asano, G.~Ishiki, S.~Shimasaki and S.~Terashima,
Phys. Rev. D \textbf{96}, no.12, 126003 (2017)
doi:10.1103/PhysRevD.96.126003
[arXiv:1701.07140 [hep-th]].



\bibitem{Hartnoll:2024csr}
S.~A.~Hartnoll and J.~Liu,
JHEP \textbf{03}, 060 (2025).

\bibitem{Komatsu:2024ydh}
S.~Komatsu, A.~Martina, J.~Penedones, A.~Vuignier and X.~Zhao,
JHEP \textbf{12}, 030 (2025).

\bibitem{Hartnoll:2025ecj}
S.~A.~Hartnoll and J.~Liu,
SciPost Phys. \textbf{19}, no.4, 099 (2025).


\end{thebibliography}
\end{document}